\documentclass{aa}
\usepackage[varg]{txfonts}
\usepackage{natbib, amsmath, amssymb, amsfonts, graphicx,xcolor}
\usepackage{mathtools,multirow}
\usepackage[citecolor=blue, linkcolor=blue, urlcolor = black, colorlinks = true]{hyperref}

\title{The near-core rotation of HD\,112429:\\a $\gamma$\,Doradus star with TESS photometry and legacy spectroscopy}

\author{
T.~Van~Reeth \inst{1}    
\and P.~De~Cat \inst{2}
\and J.~Van Beeck \inst{1,3} 
\and V.~Prat \inst{4}
\and D.~J.~Wright \inst{5} 
\and H.~Lehmann \inst{6}
\and A.-N.~Chen\'e \inst{7}
\and E.~Kambe \inst{8}
\and S.~L.~S.~Yang \inst{9}  
\and G.~Gentile \inst{10,11}
\and M.~Joos \inst{12}
}

\institute{Institute of Astronomy, KU Leuven, Celestijnenlaan 200D, 3001 Leuven, Belgium\\
           \email{timothy.vanreeth@kuleuven.be}
\and Royal Observatory of Belgium, Ringlaan 3, 1180 Brussels, Belgium
\and TAPIR, Mailcode 350-17, California Institute of Technology, Pasadena, CA 91125, USA
\and D\'epartement d'Astrophysique-AIM, CEA/DRF/IRFU, CNRS/INSU, Universit\'e Paris-Saclay, Universit\'e Paris-Diderot,Universit\'e de Paris, 91191 Gif-sur-Yvette, France
\and Centre for Astrophysics, University of Southern Queensland, West Street, Toowoomba, QLD 4350, Australia
\and Th\"uringer  Landessternwarte Tautenburg, Sternwarte 5, 07778 Tautenburg, Germany
\and Gemini Observatory/NSF's NOIRLab, 670 N. A'ohoku Place, Hilo, Hawai'i, 96720, USA
\and Subaru Telescope, National Astronomical Observatory of Japan, 650 North A'ohoku Pl., Hilo, HI, 96720 USA
\and Physics \& Astronomy Department, University of Victoria, PO Box 3055 STN CSC, Victoria, BC, V8W 3P6, Canada
\and Sterrenkundig Observatorium, Universiteit Gent, Krijgslaan 281-S9, 9000, Gent, Belgium
\and Department of Physics and Astrophysics, Vrije Universiteit Brussel, Pleinlaan 2, 1050 Brussels, Belgium
\and CEA, DAM, DIF F-91297 Arpajon, France
}

\titlerunning{Revisiting HD\,112429 with TESS}
\authorrunning{Van Reeth et al.}

\date{Received; accepted}
 
\abstract
{The TESS space mission provides us with high-precision photometric observations of large numbers of bright stars over more than 70\% of the entire sky, allowing us to revisit and characterise well-known stars.}{We aim to conduct an asteroseismic analysis of the $\gamma$\,Doradus ($\gamma$\,Dor) star HD\,112429 using both the available ground-based spectroscopy and TESS photometry, and assess the conditions required to measure the near-core rotation rate and buoyancy travel time.}{We collect and reduce the available five sectors of short-cadence TESS photometry of this star, as well as 672 legacy observations from six medium- to high-resolution ground-based spectrographs. We determine the stellar pulsation frequencies from both data sets using iterative prewhitening, do asymptotic $g$~mode modelling of the star and investigate the corresponding spectral line profile variations using the pixel-by-pixel method.}{We validate the pulsation frequencies from the TESS data up to $S/N \geq 5.6$, confirming recent reports in the literature that the classical criterion $S/N \geq 4$ does not suffice for space-based observations. We identify the pulsations as prograde dipole $g$~modes and r-mode pulsations, and measure a near-core rotation rate of $1.536\,(3)\,\rm d^{-1}$ and a buoyancy travel time $\Pi_0$ of 4190\,(50)\,s. These results are in agreement with the observed spectral line profile variations, which were qualitatively evaluated using a newly developed toy model. We establish a set of conditions that have to be fulfilled for an asymptotic asteroseismic analysis of $g$-mode pulsators. In the case of HD\,112429, two TESS sectors of space photometry suffice.}{Although a detailed asteroseismic modelling analysis is not viable for $g$-mode pulsators with only short or sparse light curves of space photometry, it is possible to determine global asteroseismic quantities for a subset of these stars. Thanks to the ongoing TESS mission, this will allow us to characterise many more stars than only those with years of data.}


\begin{document} 
\maketitle
%

\section{Introduction}
Over the last two decades, several photometric space missions such as CoRoT \citep{Auvergne2009}, {\em Kepler} \citep{Koch2010} and BRITE \citep{Weiss2014}, have been carried out to detect and study exoplanets, and provide near-continuous long-term observations of many different types of variable stars. One of the latest such missions, {\em TESS} \citep[{\em Transiting Exoplanet Survey Satellite};][]{Ricker2014}, observed $\sim70\%$ of the sky over the course of its nominal two-year mission, divided in sectors of $24\times90\,\rm deg^2$, each of which was observed for 27 days. {\em TESS} collected Full Frame Images (FFIs) at a 30-min~cadence \citep{Huang2020a,Huang2020b} and data of approximately 200\,000 preselected stars at a 2~min cadence \citep[e.g.][]{Stassun2018}. From the FFI data, we can obtain light curves for $\sim$ 10\,million moderately to very bright stars, with {\em TESS} magnitude $T \leq 13.5$ \citep{Huang2020a,Huang2020b}. In the ongoing extended mission, the TESS~FFIs are taken at a higher 10~min cadence, and per sector about 1000 stars are observed at a 20-sec. cadence \citep{Huber2021}. These large quantities of data not only allow for the detection of stellar variability in previously unstudied stars, but also enable us to do detailed follow-up studies of known variable stars.

Space-based photometry has proven to be particularly useful for the study of stars with gravito-inertial (g) mode pulsations. $G$-mode pulsations have buoyancy as the dominant restoring force and in the case of moderate to fast-rotating stars, the Coriolis force contributes as well. Gravity~modes have been detected in intermediate-mass main-sequence stars \citep[e.g.][]{Aerts2021}, the so-called $\gamma$\,Doradus \citep[$\gamma$\,Dor, with $1.4\,M_\odot \lesssim M_* \lesssim 1.8\,M_\odot$;][]{Kaye1999} and slowly-pulsating B~type \citep[SPB, with $2.5\,M_\odot \lesssim M_* \lesssim 8\,M_\odot$;][]{Waelkens1991} stars. They are most sensitive to the properties of the radiative stellar interior, just outside of the convective core. In a non-rotating, chemically homogeneous, non-magnetic star, $g$~modes with consecutive radial order $n$ but the same mode identification ($k,m$), where $k$ is the meridional degree and $m$ the azimuthal order \citep[e.g.][]{Lee1997}, are equidistantly spaced in the period domain for a pulsation frequency $\nu \ll N/2\pi$, where $N$ is the Brunt-V\"ais\"al\"a frequency \citep{Shibahashi1979,Tassoul1980}. When there is strong chemical stratification in the near-core radiative region \citep{Miglio2008a}, a magnetic field \citep[e.g.][]{Prat2019,VanBeeck2020} or non-linear coupling between the $g$~modes \citep[e.g.][]{Ouazzani2020,VanBeeck2021}, the period spacings are no longer constant, but have structure. If the star is rotating, the $g$-mode frequencies are shifted by the Coriolis force and the change between the co-rotating and the observer's reference frames \citep{Bouabid2013}. In the case of prograde ($m>0$) and zonal ($m=0$) modes, the observed period spacings decrease with increasing pulsation period. In the case of retrograde ($m<0$) $g$~modes, the period spacings mostly increase. Over the last decade, these different characteristics have been detected in hundreds of stars \citep[e.g.][]{Kurtz2014,Bedding2015,Li2020,Pedersen2021,Szewczuk2021,Garcia2022} and analysed to measure the near-core rotation rates \citep[e.g.][]{VanReeth2016,Christophe2018,Li2020,Takata2020b,Takata2020,Szewczuk2021}, probe the core boundary mixing \citep[CBM; e.g.][]{Michielsen2021,Mombarg2021,Pedersen2021}, opacity \citep[e.g.][]{Szewczuk2018,Walczak2019} and mixing processes in the radiative envelope \citep[e.g.][]{Mombarg2020,Pedersen2021,Mombarg2022} and constrain internal magnetic fields \citep[e.g.][]{Buysschaert2018,Prat2019}. Since recently, the study of $g$~modes has also revealed information about the size \citep[e.g.][]{Johnston2021,Mombarg2021,Pedersen2021}, rotation rate \citep{Ouazzani2020,Lee2021,Saio2021} and other properties \citep{Aerts2021b} of the convective core itself.

Because $g$-mode pulsations have low amplitudes (often $\lesssim$ 1\,mmag) and periods between 0.3 and 3 days, they are hard to detect from ground-based observations alone. By contrast, space photometry reaches a much higher signal-to-noise ratio ($S/N$) and duty cycle (often up to $\gtrsim 90\,\%$). However, because the density of $g$-mode pulsation frequency spectra is very high, with spacings between consecutive pulsation frequencies on the order of 0.001 to 0.01\,$\rm d^{-1}$, near-continuous observations with a time span of years are often needed to resolve individual modes. This condition is not fulfilled for most stars observed with {\em TESS}, except for those located in one of the {\em TESS} continuous viewing zones (CVZs). On the other hand, ground-based observations of known $g$-mode pulsators can cover longer time spans.

In this work, we revisit the well-studied $\gamma$\,Dor star HD\,112429. It was first reported as a $\gamma$\,Dor candidate by both \citet{Aerts1998} and \citet{Handler1999} based on analyses of the Hipparcos photometry and subsequently included in several follow-up campaigns. \citet{Henry2005} confirmed these findings and measured five pulsation frequencies from 460 photometric observations spanning 233 days. \citet{Fekel2003} obtained two coud\'e spectra and found it to be a fast-rotating single star with spectral type F1. These results were confirmed by \citet{Mathias2004}, who took 40 spectra over a time span of 566 days. The authors observed that the line profile variability was concentrated in the wings of the spectral lines, and recovered the dominant $g$-mode frequency from a frequency analysis. More recently, \citet{Kahraman2016} measured the atmospheric parameters from the Fe~lines in a spectroscopic analysis (listed in Table\,\ref{tab:hd112429}). Finally, a cold debris disk has been detected around the star based on observations with the Spitzer space telescope \citep[e.g.][]{Chen2005,Plavchan2009}. Consequently, HD\,112429 has been included in sample studies to look for exoplanets \citep[e.g.][]{Janson2013}, but so far none have been detected.

We conducted a combined asteroseismic analysis of the available {\em TESS} photometry and legacy ground-based spectra of HD\,112429, which were taken as part of the multi-site observation campaign outlined by \citet{DeCat2009}. Here, the atmospheric parameters by \citet{Kahraman2016}, listed in Table\,\ref{tab:hd112429}, place sufficiently strong and accurate constraints on HD\,112429 and were therefore used to guide our analysis. 

In Sect.\,\ref{sec:obs} we describe the general characteristics of the different data sets and the data reduction process. In Sect.\,\ref{sec:methods-analysis} we determined the pulsation frequencies by iteratively prewhitening both the spectroscopic and photometric data sets. Next, in Sect.\,\ref{sec:core-rot}, we used these frequencies to determine the near-core rotation rate and the buoyancy travel time of the star, and evaluated the sampling bias introduced by the sparse TESS sectors during which HD\,112429 was observed. In Sect.\,\ref{sec:lpv} we analyse the spectroscopic line profile variations and compared the results with those obtained in Sect.\,\ref{sec:core-rot}. Finally, we present our conclusions in Sect.\,\ref{sec:discussion}.

\begin{table}
\caption{\label{tab:hd112429} Atmospheric parameter values of HD\,112429, collected from \citet{Kahraman2016}.}
\centering
\begin{tabular}{lc}
\hline\hline
$T_{\rm eff}$ (K) & 7200 $\pm$ 100\\
$\log\,g$ (dex) & 3.9 $\pm$ 0.2\\
$v_{\rm micro}$ (km\,$\rm s^{-1}$) & 3.0 $\pm$ 0.2\\
$v\sin i$ (km\,$\rm s^{-1}$) & 120 $\pm$ 3\\
$\log\,\epsilon_{Fe}$ & 7.29 $\pm$ 0.23\\
\hline
\end{tabular}
\end{table}

\section{Observations}
\label{sec:obs}
\subsection{TESS photometry}
\label{subsec:TESS-data}
HD\,112429 has been observed by TESS in short-cadence during sectors 14, 15, 21 and 22 in cycle 2, and sector 41 in cycle 4, spanning a total time of 763 days. To ensure that the $S/N$ of our light curves are maximal and the instrumental trends minimal, we re-extracted the light curves from the target pixel files (TPF) with optimised aperture masks, instead of using the simple aperture photometry (SAP) or the Pre-search Data Conditioned data (PDC-SAP) provided by the TESS-SPOC pipeline \citep{Jenkins2016}. 

For each TPF, a suitable binary aperture mask was determined based on the median frame of the cutout. We required that\begin{itemize}
    \item[{\em (i)}] selected pixels have a flux count larger than a $3-\sigma$ threshold above the median flux of this frame, whereby the $\sigma$-value is calculated as $$\sigma = 1.4826\overline{|f_{p,i} - \overline{f_p}|},$$ with $\overline{X} = {\rm median}(X_i)$ and $f_{p,i}$ the flux of the $i^{th}$ pixel. Here, the value 1.4826 is the scaling factor between the standard deviation $\sigma$ and the median absolute deviation \citep{Rousseeuw1993}. For this requirement, we use the aperture mask selection defined in \texttt{lightkurve}\footnote{ \href{https://docs.lightkurve.org/}{https://docs.lightkurve.org/}} \citep{Lightkurve}.
    \item[{\em (ii)}] there is only one local maximal flux count included within the mask.
    \item[{\em (iii)}] none of the selected pixels are included in other aperture masks, constructed for nearby bright stars.
\end{itemize}
For bright stars such as HD\,112429, which has an apparent magnitude $m_V$ of 5.23\,mag, stellar flux is sometimes lost via bleeding columns on the CCD when the star is at its brightest. Hence, the obtained aperture masks were inspected visually for the individual frames within each TPF.

The local background flux count per pixel $b_{p}(t)$, caused by sunlight that is reflected by the Earth and enters the telescope, was accurately estimated by the SPOC pipeline. It was validated by a comparison with the $5^{th}$-percentile flux count for each frame within the TPF. We ensured that the background flux estimates did not contain any contaminating signal from nearby variable stars using visual inspection.

The extracted light curve was then calculated and converted to mmag by taking $$f(t) = -2500{}^{10}\log\left(\overline{\sum_i \left[f_{p,i}(t) - b_p(t)\right]}\right),$$ and it was normalised by subtracting the mean. Finally, time stamps with potentially bad data points were identified by means of $5-\sigma$ clipping per pixel in the aperture mask. These data points were then inspected visually in the reduced light curve, and removed if necessary.

\begin{figure*}
    \centering
    \includegraphics[width=\textwidth]{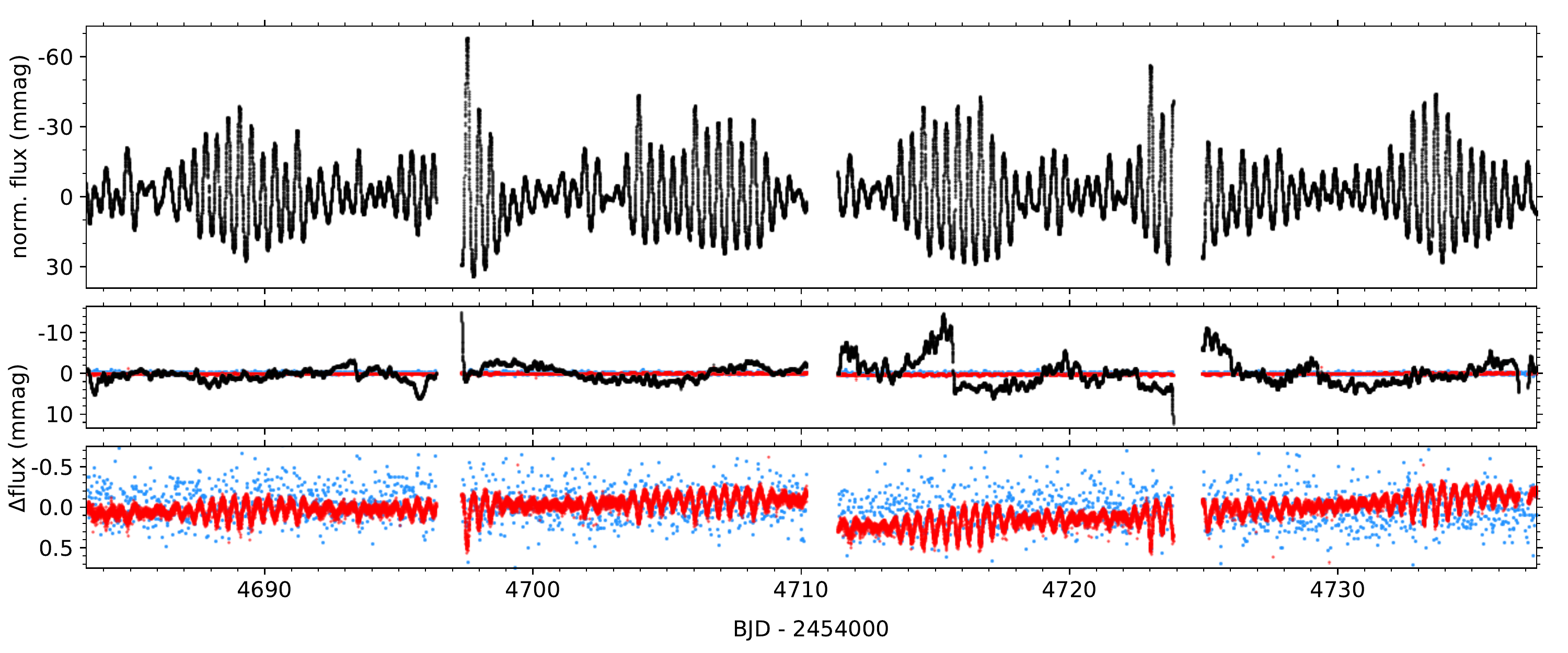}
    \caption{ {\em Top:} custom reduced short-cadence TESS light curve of HD\,112429, illustrated for sectors 14 and 15. The gaps in the the light curve are caused by the data downlinks of TESS. {\em Middle:} absolute differences between our custom reduced light curve and the FFI~lightcurve (blue), TESS-SPOC SAP flux (red) and PDC-SAP flux (black). {\em Bottom:} zoom-in of absolute differences between our custom reduced light curve and the FFI~lightcurve (blue) and TESS-SPOC SAP flux (red).}
    \label{fig:tess-lc}
\end{figure*}

To validate the final reduced light curve, we then compared it to the standard SAP and PDC-SAP light curves, as well as custom light curves extracted from the FFI \citep[using CCD cutouts of $25\times25$ pixels, obtained with \texttt{TessCut};][]{Brasseur2019}. This is illustrated in Fig.\,\ref{fig:tess-lc}. The PDC-SAP flux exhibits strong instrumental variability, but no notable differences were found between the normalised SAP flux and the re-extracted short-cadence and FFI light curves. However, as can be seen in the bottom panel of Fig.\,\ref{fig:tess-lc}, the $S/N$ of the re-extracted short-cadence light curve is slightly higher than the $S/N$ of the SAP and FFI light curves. The additional noise of the SAP light curve originates from the smaller aperture mask, whereas the added noise in the FFI light curve is caused by the different background flux estimation, which for the FFI is calculated as the median $e^-\,s^{-1}$ count of the surrounding pixels in the CCD that do not contain significant stellar flux. Hence, the re-extracted short-cadence light curve will be used during the rest of this work.

\subsection{Spectroscopy}
\label{subsec:spectroscopy}
\subsubsection{Data collection}
\label{subsubsec:spectra-collection}
To complement the TESS~photometry, we have retrieved 672 legacy observations from the spectroscopic multi-site campaign outlined by \citet{DeCat2009}. These were taken with six medium- to high-resolution spectrographs, as summarised in Table \ref{tab:observations}: the HERMES spectrograph (at the 1.2-m Mercator telescope, Observatorio del Roque de los Muchachos, La Palma, Spain), the HIDES spectrograph (at the 1.88-m telescope at Okayama Astrophysical Observatory, Japan), the McKellar spectrograph (at the 1.2-m telescope at the Dominion Astrophysical Observatory, Victoria, Canada), the SES spectrograph (at the 2.1-m telescope at the McDonald Observatory in Texas, USA), the SOPHIE spectrograph (at the 1.93-m telescope at the Observatoire de Haute-Provence, France) and the TCES spectrograph (at the 2.0-m Alfred Jensch Telescope at the T\"uringer Landessternwarte in Tautenburg, Germany). The McKellar\footnote{\href{https://www.cadc-ccda.hia-iha.nrc-cnrc.gc.ca/en/search/}{https://www.cadc-ccda.hia-iha.nrc-cnrc.gc.ca/en/search/}}, HIDES\footnote{\href{https://smoka.nao.ac.jp/index.jsp}{https://smoka.nao.ac.jp/index.jsp}} and SOPHIE\footnote{\href{http://atlas.obs-hp.fr/sophie/}{http://atlas.obs-hp.fr/sophie/}} observations are nowadays freely available from their respective public archives.

When possible, such as for HERMES and SOPHIE, we made use of the available standard data reduction pipelines for the different instruments. The TCES and SES spectra have been reduced using standard ESO-MIDAS packages\footnote{ \href{https://www.eso.org/sci/software/esomidas/}{https://www.eso.org/sci/software/esomidas/}} while the HIDES and McKellar spectra were subjected to the normal reduction procedures using IRAF\footnote{ \href{https://github.com/iraf-community/iraf}{https://github.com/iraf-community/iraf}; IRAF is distributed by the National Optical Astronomy Observatories, which is operated by the Association of Universities for Research in Astronomy, Inc. under cooperative agreement with the National Science Foundation, USA.} The reduction included bias subtraction, filtering of cosmic ray events, flat fielding using Halogen lamps, order extraction, wavelength calibration using a ThAr~lamp, and normalisation and merging of spectral orders. A large number of telluric $O_2$-lines was used to correct for small instrumental shifts.

\begin{table*}
	\centering
	\caption{Overview of the collected observations that were used in our analysis, listed per spectrograph. For each observational campaign, the time, spectral resolution, covered wavelength range and average $\langle S/N\rangle $ are also given. Here, the $\langle S/N \rangle$ were calculated in a $1$\,nm-margin around wavelength $\lambda  = 550\,$nm, except for the McKellar spectra, where the whole available wavelength range was used. For each spectrum, the $S/N$ value was estimated as the inverse relative scatter with respect to the scaled average spectrum.}
	\label{tab:observations}
	\begin{tabular}{llrrrrr} 
		\hline\hline
		spectrograph & observatory                             & time & $R$     & $\lambda$ (nm) & \# obs. & $\langle S/N \rangle$\\
		\hline
		HERMES${}^{(1)}$       & 1.2-m Mercator telescope,      & Mar-May 2009 &  85\,000 & 377-900 &  86 & 181\\
		                       & \qquad Roque de los Muchachos, Spain  & Jan 2010   &  85\,000 & 377-900 &  82 & 294\\ 
		HIDES${}^{(2,3)}$      & 1.88-m telescope, & Dec 2008 & 50\,000 & 386-774 & 7 & 222\\
		                       & \qquad Okayama Astrophysical Observatory, Japan                                                      & Mar 2009      & 50\,000 & 386-774 & 165 & 205\\
		                       &                                                       & Jul-Aug 2009  & 50\,000 & 386-774 & 19 & 150\\ 
		McKellar${}^{(4)}$     & 1.2-m telescope, & May-Nov 2009 &  60\,000 & 445-460 &  34 & 157\\
		                       & \qquad Dominion Astrophysical Observatory, Canada & & & & \\
		SES${}^{(5)}$          & 2.1-m telescope,    & Dec 2008 &  60\,000 & 543-668 &  31 & 195\\
				               & \qquad McDonald Observatory, Texas, USA   & Jun 2009     &  60\,000 & 518-620 &  32 & 195\\
		SOPHIE${}^{(6,7)}$     & 1.93-m telescope, &  Jan 2008 & 76\,000  & 387-694 &  57 & 246\\
		                       & \qquad Observatoire de Haute-Provence, France   & Dec 2008 & 40\,000  & 387-694 &   3 & 216\\
		                       &                                                        & Jul 2009     & 40\,000  & 387-694 &  90 & 220\\
		TCES                   & 2.0-m telescope,  & Jul-Oct 2009  &  67\,000 & 472-736 &                        66 & 173\\
		                       & \qquad T\"uringer Landessternwarte, Germany & & & & \\
		\hline
	\end{tabular}
\tablebib{
(1)~\citet{Raskin2011}; (2)~\citet{Izumiura1999}; (3)~\citet{Kambe2002}; (4)~\citet{Monin2014}; (5)~\citet{McCarthy1993}; (6)~\citet{Perruchot2008}; (7)~\citet{Bouchy2009}
}
\end{table*}

\subsubsection{Normalisation and order merging}
\label{subsubsec:normalisation}

Both the format and quality of the reduced data products differ strongly between spectrographs. To ensure a homogeneous analysis, the different spectra were carefully evaluated and renormalised, barycentrically corrected where needed, and the different spectral orders merged when this step was not included in the preceding data reduction.

We first collected and normalised 181 reduced HERMES spectra by fitting a third-order polynomial spline to carefully selected continuum points, and removed cosmic hits by applying 5-$\sigma$ clipping. To ensure that possible residuals from the instrument response function do not affect our analysis of the spectral line profile variability, the same wavelength knot points were chosen for each HERMES spectrum to construct the continuum spline model. Following the normalisation, we visually inspected the data and excluded thirteen spectra for which the spectral order merging had failed. We then applied the barycentric correction to the 86 spectra taken from March to May 2009. For the 82 observations taken in 2010, this correction was already included in the data reduction. Finally, we calculated the average spectrum of the accepted 168 HERMES observations, which then served as a reference spectrum to normalise the observations from the other five spectrographs.

Next we focused on the McKellar and the TCES spectroscopy, which did not require spectral order merging either. While the McKellar spectra each consist of a single order, the preceding TCES data reduction already included reliable order merging. To ensure that the normalised McKellar and TCES spectra matched the normalised HERMES spectra as closely as possible, we first applied the required barycentric corrections, divided the reduced observations by the reference HERMES spectrum and fitted third-order polynomial splines to the resulting fractions. This minimised the influence of strong metal and hydrogen lines on the fitting of the different response functions. And we again used a common series of knot points for each spectrograph. The resulting spline functions were then used to normalise the observed spectra.

\begin{figure}
    \centering
    \includegraphics[width=88mm]{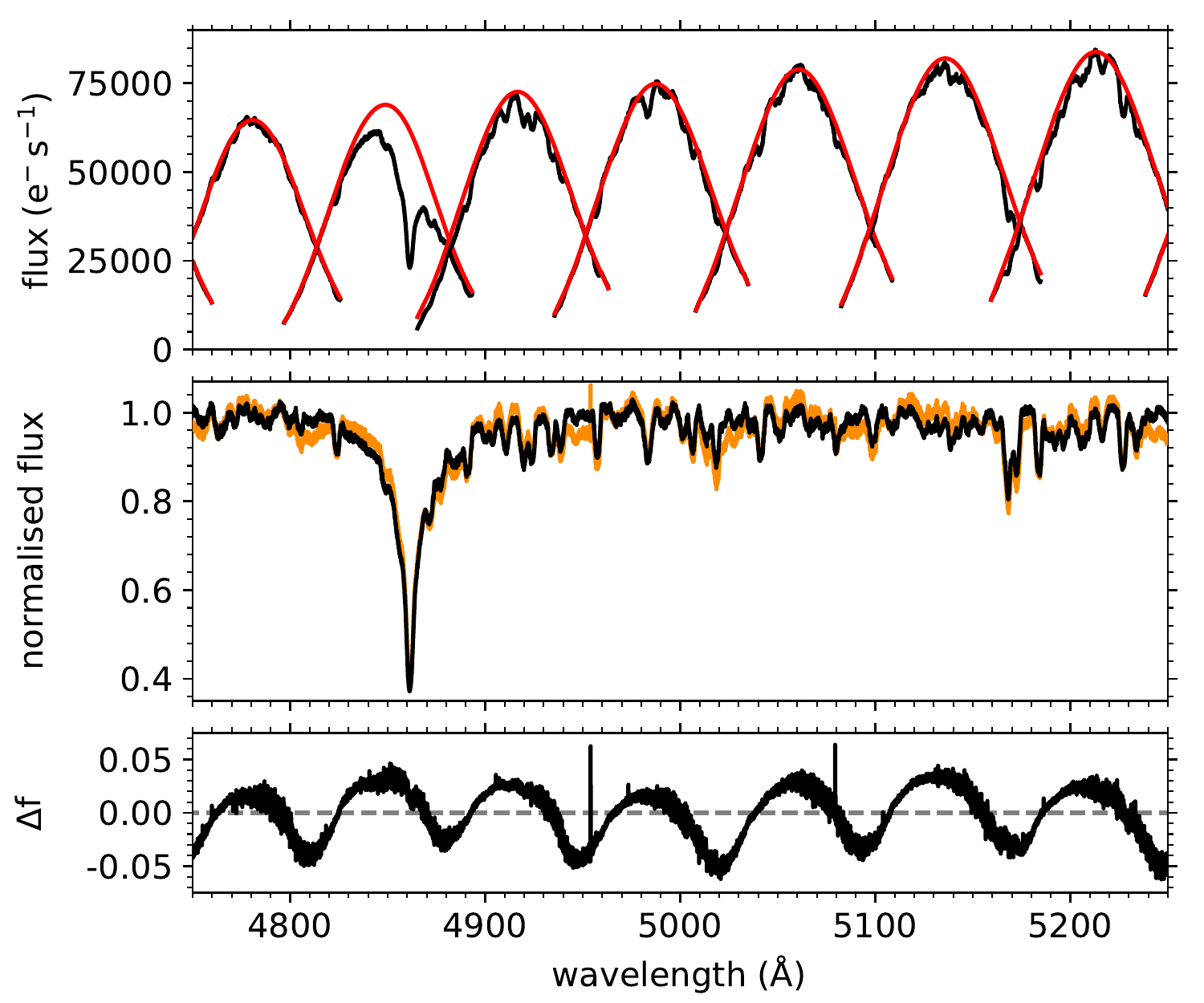}
    \caption{The normalisation of the SOPHIE spectrum taken at BJD 2454828.69749. {\em Top:} the pipeline-reduced individual spectral orders (black) and the best-fitting scaled blaze functions (red) defined in Eqs.\,(\ref{eq:blaze-sc}) and (\ref{eq:blaze}). {\em Middle:} our normalised and merged spectrum (black) compared with the normalised one-dimensional spectrum from the SOPHIE pipeline (orange). {\em Bottom:} the difference between the two normalised spectra shown in the middle panel. The residual instrumental trends from the one-dimensional pipeline-reduced spectrum dominate.}
    \label{fig:sophie-spec}
\end{figure}

Finally, we normalised the remaining non-merged spectra from the SES, HIDES and SOPHIE spectrographs. With the exception of SOPHIE observations, only separate spectral orders were available for these reduced spectra. For the SOPHIE spectrograph, there were residual discontinuities in the merged spectra which affected our subsequent analysis. To normalise each spectrum, we barycentrically corrected the data, divided every spectral order by the reference HERMES spectrum, and fitted a scaled semi-analytical blaze function $B_{sc}(\lambda)$ to the result of this division:
\begin{equation}
    B_{sc}(\lambda) = \sum_{i=0}^n \sum_{j=0}^{n-i} c_{ij} B(\lambda)^i\left(\lambda-\lambda_c\right)^j,\label{eq:blaze-sc}
\end{equation}
where the Blaze function $B(\lambda)$ \citep{Barker1984} is given by
\begin{equation}
B(\lambda) = \frac{\sin^2\left(\pi\alpha\left(\lambda-\lambda_c\right)\right)}{\left(\pi\alpha\left(\lambda-\lambda_c\right)\right)^2}.\label{eq:blaze}
\end{equation} 
Here $\lambda$ is the wavelength, $\lambda_c$ is the central wavelength of the evaluated spectral order, and the coefficients $c_{ij}$ and $\alpha$ are scaling factors that have to be optimised. We took the degree $n=3$, and evaluated the relative contribution of each term using a Bayesian Information Criterion (BIC) \citep[e.g.][]{KassRaftery1995}. This allowed us to account for the instrument-dependent characteristics, and by modelling the ratio between the observed spectra and the reference HERMES spectrum, we avoided overfitting the Balmer lines and strong metal lines in the observed spectra. This is illustrated in the top panel of Fig.\,\ref{fig:sophie-spec} for the SOPHIE spectrum taken at BJD 2454828.69749. Following the normalisation of the individual spectral orders, we merged them by linearly scaling and adding the flux of the overlapping parts between consecutive orders. Finally, we applied additional cosmic clipping (with a 5-$\sigma$ limit) to all spectra. This is illustrated in the middle panel of Fig.\,\ref{fig:sophie-spec}, where we compared our normalised and merged spectrum with the corresponding one-dimensional spectrum from the SOPHIE pipeline, which we renormalised with a $\rm 3^{rd}$-order polynomial spline. Both the quality and necessity of our own spectrum normalisation and order merging can clearly be seen in the bottom panel, where the residual instrumental trends of the full pipeline-reduced spectrum are shown.

Because our data reduction relied heavily on the comparison between each spectrum and the average HERMES spectrum after applying barycentric correction, it failed in wavelength regions that are dominated by telluric lines, such as between 627.5\,nm and 632.5\,nm. Hence, in the remainder of this work, we avoid those regions.

\subsubsection{Least-squares deconvolution profiles}
\label{subsubsec:lsd-profiles}
The spectra from the different instruments cover very different wavelength ranges. In order to analyse all available data consistently and maximise the signal-to-noise ratio of the data, we calculated the least-squares deconvolution (LSD) profiles of the spectra \citep[e.g.][]{Donati1997,Koch2010} using the code developed by \citet{Tkachenko2013}\footnote{\href{https://github.com/TVanReeth/Least-squares-deconvolution}{https://github.com/TVanReeth/Least-squares-deconvolution}}. These are effectively high-quality, high-$S/N$ average spectral lines, calculated by convolving the observed spectra with a spectral line mask and accounting for the blending of overlapping lines. Here we used a line mask calculated following the approach from \citet{Tkachenko2013}, for the atmospheric parameter values listed in Table\,\ref{tab:hd112429}. The spectral line strengths were computed with the SynthV code \citep{Tsymbal1996}, using information from the Vienna Atomic Line Database \citep[VALD;][]{Kupka1999} and atmosphere models from the LLmodels code \citep{Shulyak2004}. The mask covered a wavelength range from 440\,nm\ to 480\,nm\ and from 492\,nm\ to 580\,nm. Hence, we included the dominant metal line regions within the observed spectra and avoided contamination from Balmer lines and from telluric lines. All spectral lines with strengths of 0.01\,\%, that is 17\,945 lines in total, are included in the complete line mask.

\begin{figure}
    \centering
    \includegraphics[width=88mm]{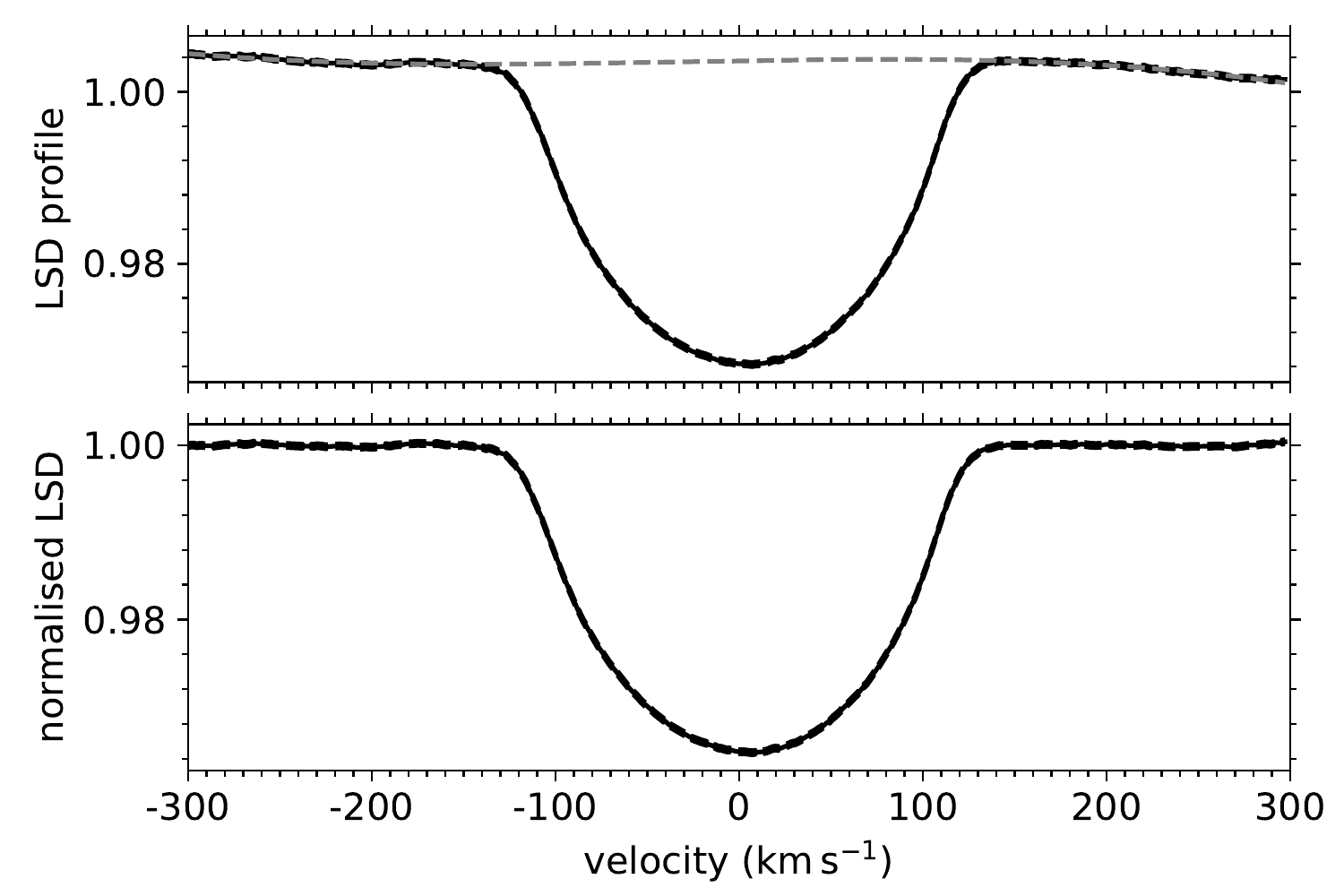}
    \caption{{\em Top:} Average LSD profile of the HERMES spectra (full black line) with its error margins (black dashed line). The dashed grey line indicates the $\rm 3^{rd}$-order polynomial used to renormalise the LSD~profile. {\em Bottom:} The resulting renormalised LSD~profile.}
    \label{fig:lsd-profiles}
\end{figure}

\begin{figure*}
    \centering
    \includegraphics[width=\textwidth]{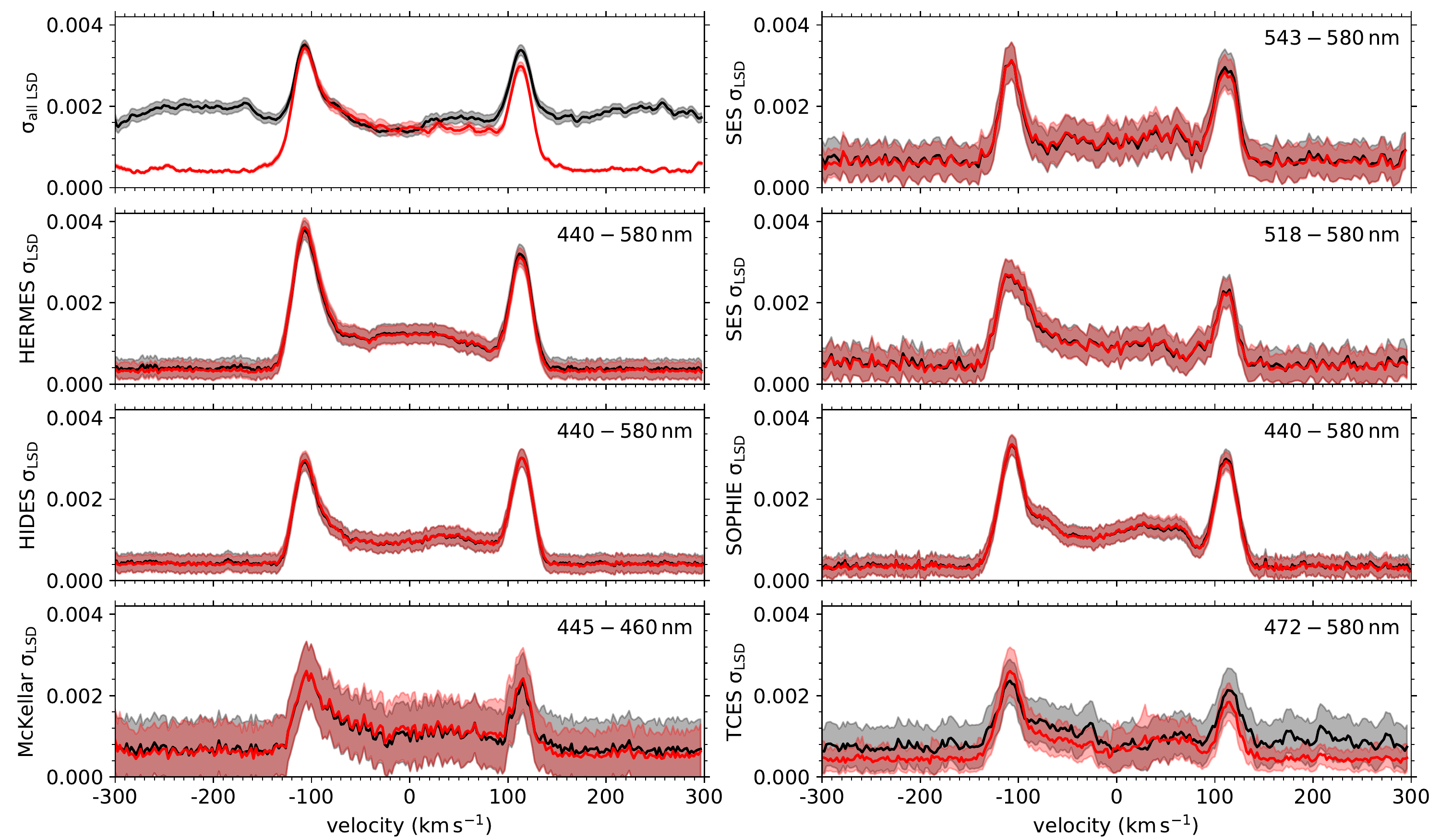}
    \caption{Standard deviations $\sigma$ for the LSD~profiles, both for the entire data set (top left panel) and per spectrograph (other panels), calculated before (black) and after (red) the renormalisation of the LSD~profiles. The coloured areas indicate the 1-$\sigma$~uncertainties, which were calculated with Monte Carlo simulations. The wavelength ranges used for the LSD calculations are given in the top right corners.}
    \label{fig:lsd-prof-variations}
\end{figure*}

As an example the mean of the HERMES LSD profiles is shown in Fig.\,\ref{fig:lsd-profiles} and as can be seen, the LSD~profiles do not have a continuum flux at unity. This is because spectral line blending is partially non-linear (for example, caused by pressure broadening of the spectral lines), while it is assumed to be linear in the LSD algorithm \citep{Tkachenko2013}. This produces artefacts in the LSD continuum on the scale of the profile width, that depend on the used spectrograph and on the spectral wavelength range used for the LSD~calculations. To remove these artefacts without affecting the intrinsic line profile variations, we renormalised the LSD profile by fitting a $\rm 3^{rd}$-order polynomial to the LSD continuum flux, as demonstrated in Fig.\,\ref{fig:lsd-profiles}. And as illustrated in Fig.\,\ref{fig:lsd-prof-variations}, this allowed us to combine the LSD~profiles from the different spectrographs in an analysis, which is discussed further in Sect.\,\ref{sec:methods-analysis}.

\section{Frequency analysis}
\label{sec:methods-analysis}
\begin{figure*}
    \centering
    \includegraphics[width=\textwidth]{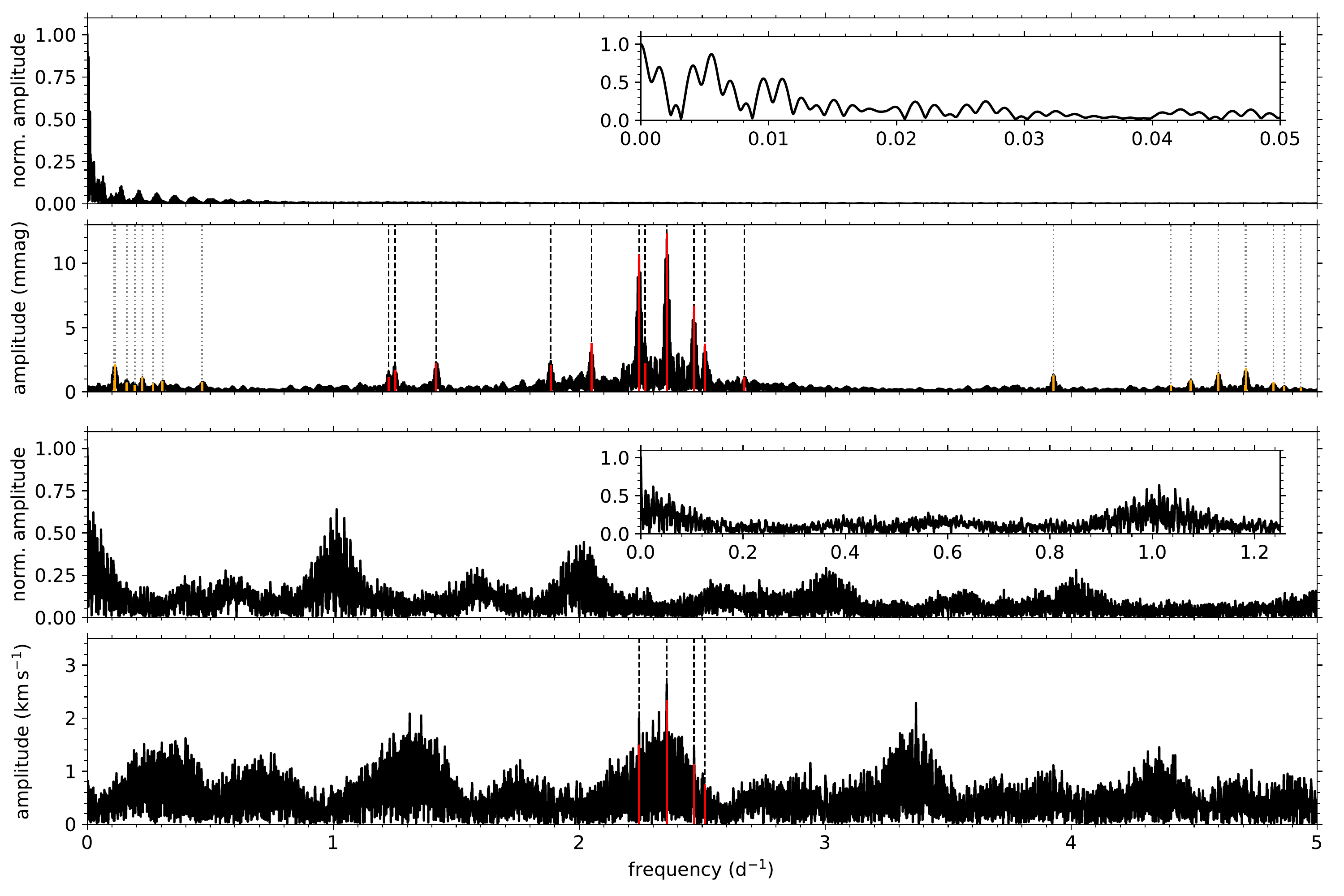}
    \caption{Lomb-Scargle periodograms of the full TESS light curve ({\em top two panels}) and the $1^{\rm st}$ moments calculated from the LSD~profiles ({\em bottom two panels}), with the spectral windows of both data sets. {\em $1^{\rm st}$ panel:} spectral window of the full TESS light curve, with a zoom-in shown in the inset axes. {\em $2^{\rm nd}$ panel:} Lomb-Scargle periodogram of the full TESS light curve (black) with the corresponding prewhitened pulsation frequencies with $S/N \geq 5.6$. The red and dashed black lines mark independent $g$-mode pulsations. The orange and dotted lines indicate the combination frequencies. {\em $3^{\rm rd}$ panel:} spectral window of the $1^{\rm st}$ moments, with a zoom-in shown in the inset axes. {\em $4^{\rm th}$ panel:} Lomb-Scargle periodogram of the $1^{\rm st}$ moments (black) with the corresponding prewhitened pulsation frequencies with $S/N \geq 4.0$. The red and dashed black lines mark independent $g$-mode pulsations.}
    \label{fig:puls-freq}
\end{figure*}

Following the detailed data reduction, we first focused on the variability analysis of the reduced TESS photometry. Here, we combined the light curves from the five available sectors and applied iterative prewhitening with the code used by \citet{VanBeeck2021}. The $g$-mode pulsation frequencies were determined by non-linearly fitting sine waves to the data, that is, by optimising the frequencies, amplitudes and phases of these waves. This was done in order of decreasing $S/N$ values, which were calculated by taking the ratios between the amplitude of the relevant peak in the Lomb-Scargle periodogram \citep{Scargle1982} and the mean signal in the surrounding $1\,\rm d^{-1}$-window. The pulsation frequencies were accepted when $S/N \geq 4.0$ \citep{Breger1993}. However, following the methodology by \citet{Zong2016}, we confirm that a cutoff value of $S/N \geq 5.6$ is required to have $\lesssim 0.01\%$ chance of measuring a noise frequency. The final measured pulsation frequencies are illustrated in Fig.\,\ref{fig:puls-freq} and listed in Table\,\ref{tab:tess-freq}. For completeness, all pulsation frequencies with $S/N \geq 4.0$ are included, but frequencies with $S/N \geq 5.6$ are marked with ``${}^*$''.

Second, we determined the pulsation frequencies from the LSD profile variations. To this end, we calculated the radial velocity variations using centroids \citep[e.g.][]{Aerts1992} and again applied iterative prewhitening. Pulsation frequencies were accepted when $S/N > 4$ \citep{Breger1993}. However, despite the normalisation corrections that we applied to the LSD~profiles in the previous Section, small-scale differences in the velocity calibration of the different spectrographs remained present in the data. Hence, when we fitted sine waves to the data, we accounted for these radial velocity shifts by including the zero-point $v_0$ as a free parameter per instrument. These velocity offsets are listed in Table\,\ref{tab:spec-v0}, while the measured pulsation frequencies are listed in Table\,\ref{tab:spec-freq} and illustrated in Fig.\,\ref{fig:puls-freq}. We can see that these frequencies match the four dominant pulsations measured from the TESS photometry, which are indicated in bold in Table\,\ref{tab:tess-freq}. We also note that the three spectrographs with the largest radial velocity offsets in Table\,\ref{tab:spec-v0}, McKellar, SES and TCES, correspond to mostly noisy profiles shown in Fig.\,\ref{fig:lsd-prof-variations}.

\begin{table}
\caption{\label{tab:tess-freq} Frequencies determined from the available TESS photometry using iterative prewhitening, with the corresponding amplitudes, pulsation phases and $S/N$~values. Pulsation frequencies that fulfil the $S/N \geq 5.6$ criterion from \citet{Zong2016} are additionally marked with ``${}^*$'', while the four dominant pulsations are indicated in bold. The dominant $2^{\rm nd}$-order combination frequencies are indicated in the last column.}
\centering
\begin{tabular}{lllllll}
\hline\hline
            & frequency    & ampl.  & phase       & $S/N$    & comments\\
            & (d$^{-1}$)   & (mmag)     & (rad)       &        & \\
\hline
 $\nu_{1}$    & $0.1097(1)$   & $0.9(1)$  & $-2.5(2)$    & $4.2$  & $\nu_{25}^* - \nu_{22}$\\
 $\nu_{2}^*$  & $0.11251(5)$  & $2.2(1)$  & $1.89(7)$    & $6.9$  & $\nu_{25}^* - \nu_{21}^*$\\
 $\nu_{3}$    & $0.1604(2)$   & $0.7(1)$  & $-0.4(2)$    & $4.4$  & $\nu_{29}^* - \nu_{27}^*$\\
 $\nu_{4}$    & $0.1932(2)$   & $0.6(1)$  & $-0.6(2)$    & $4.0$  & $\nu_{21}^* - \nu_{18}^*$\\
 $\nu_{5}$    & $0.22364(9)$  & $1.2(1)$  & $0.5(1)$     & $5.2$  & $\nu_{26}^* - \nu_{21}^*$\\
 $\nu_{6}$    & $0.2674(2)$   & $0.7(1)$  & $1.2(2)$     & $4.7$  & $\nu_{27}^* - \nu_{21}^*$\\
 $\nu_{7}$    & $0.3057(1)$   & $0.8(1)$  & $1.0(2)$     & $4.3$  & $\nu_{25}^* - \nu_{18}^*$\\
 $\nu_{8}$    & $0.4663(1)$   & $0.8(1)$  & $0.9(2)$     & $4.2$  & $\nu_{13}^* - \nu_{11}^*$\\
 $\nu_{9}^*$  & $1.22473(9)$  & $1.2(1)$  & $3.0(1)$     & $6.1$  & \\
 $\nu_{10}^*$   & $1.25143(6)$  & $1.7(1)$  & $2.96(9)$  & $8.4$  & \\
 $\nu_{11}^*$  & $1.41755(5)$  & $2.2(1)$  & $-2.77(7)$  & $8.7$  & \\
 $\nu_{12}$    & $1.8737(1)$   & $0.8(1)$  & $1.4(2)$    & $4.1$  & \\
 $\nu_{13}^*$  & $1.88384(5)$  & $2.1(1)$  & $-0.65(7)$  & $7.0$  & \\
 $\nu_{14}$    & $1.9332(1)$   & $0.8(1)$  & $-2.2(2)$   & $4.2$  & \\
 $\nu_{15}$    & $1.9567(1)$   & $0.7(1)$  & $-1.8(2)$   & $4.3$  & \\
 $\nu_{16}$    & $2.0001(1)$   & $0.8(1)$  & $-1.7(2)$   & $4.0$  & \\
 $\nu_{17}$    & $2.01949(8)$  & $1.4(1)$  & $-0.7(1)$   & $5.3$  & \\
 $\nu_{18}^*$   & $2.04992(3)$  & $3.8(1)$  & $-0.56(4)$ & $10.2$ & \\
 $\nu_{19}$    & $2.1318(1)$   & $0.8(1)$  & $0.4(2)$    & $4.2$  & \\
 $\nu_{20}$    & $2.19929(9)$  & $1.2(1)$  & $-1.4(1)$   & $5.2$  & \\
 $\mathbf{\nu_{21}^*}$   & $\mathbf{2.24312(1)}$  & $\mathbf{10.7(1)}$ & $\mathbf{-0.36(1)}$ & $\mathbf{13.3}$ & \\
 $\nu_{22}$    & $2.24589(6)$  & $1.5(1)$  & $-2.2(1)$   & $5.5$  & \\
 $\nu_{23}^*$  & $2.26799(5)$  & $2.2(1)$  & $-2.73(7)$  & $7.1$  & \\
 $\nu_{24}$    & $2.3240(1)$   & $0.9(1)$  & $-1.8(2)$   & $4.2$  & \\
 $\mathbf{\nu_{25}^*}$   & $\mathbf{2.355597(9)}$ & $\mathbf{12.4(1)}$ & $\mathbf{1.07(1)}$  & $\mathbf{11.6}$ & \\
 $\mathbf{\nu_{26}^*}$   & $\mathbf{2.46671(2)}$  & $\mathbf{6.7(1)}$  & $\mathbf{0.48(2)}$  & $\mathbf{12.7}$ & \\
 $\mathbf{\nu_{27}^*}$   & $\mathbf{2.51053(3)}$  & $\mathbf{3.8(1)}$  & $\mathbf{1.30(4)}$  & $\mathbf{9.6}$  & \\
 $\nu_{28}$    & $2.5793(2)$   & $0.7(1)$  & $-2.5(2)$   & $4.5$  & \\
 $\nu_{29}^*$  & $2.67077(9)$  & $1.2(1)$  & $-0.5(1)$   & $6.6$  & \\
 $\nu_{30}$    & $3.7830(2)$   & $0.5(1)$  & $2.3(2)$    & $5.0$  & \\
 $\nu_{31}^*$   & $3.92793(8)$  & $1.3(1)$  & $-1.2(1)$  & $9.7$  & $\nu_{11}^* - \nu_{27}^*$\\
 $\nu_{32}$    & $4.4055(2)$   & $0.5(1)$  & $2.3(2)$    & $5.2$  & $\nu_{18}^* + \nu_{25}^*$\\
 $\nu_{33}^*$  & $4.4862(1)$   & $0.9(1)$  & $1.1(2)$    & $6.8$  & $2\nu_{21}^*$\\
 $\nu_{34}^*$   & $4.59868(6)$  & $1.5(1)$  & $2.5(1)$   & $9.4$  & $\nu_{21}^* + \nu_{25}^*$\\
 $\nu_{35}$    & $4.7084(2)$   & $0.5(1)$  & $-0.4(2)$   & $5.3$  & \\
 $\nu_{36}^*$   & $4.71119(6)$  & $1.8(1)$  & $-2.30(8)$ & $9.7$  & $2\nu_{25}^*$\\
 $\nu_{37}^*$  & $4.8223(1)$   & $0.7(1)$  & $-2.4(2)$   & $6.8$  & $\nu_{25}^* + \nu_{26}^*$\\
 $\nu_{38}$    & $4.8661(2)$   & $0.5(1)$  & $-2.1(2)$   & $5.3$  & $\nu_{25}^* + \nu_{27}^*$\\
 $\nu_{39}$    & $4.9334(2)$   & $0.3(1)$  & $2.8(5)$    & $5.1$  & $2\nu_{26}^*$\\
\hline
\end{tabular}
\end{table}

\begin{table}
\caption{\label{tab:spec-v0} Zero-point radial velocity corrections $v_0$ for the different spectrographs, determined during the frequency analysis of the radial velocity variations.}
\centering
\begin{tabular}{lc}
\hline\hline
 spectrograph    & zero-point $v_0$    \\
                 & ($\rm km\,s^{-1}$)  \\
\hline
 HERMES   &  2.16(10)\\
 HIDES    &  1.99(9)\\
 McKellar &  -4.7(3)\\
 SES      &  4.31(17)\\
 SOPHIE   &  0.54(9)\\
 TCES     &  4.9(3)\\
\hline
\end{tabular}
\end{table}

\begin{table}
\caption{\label{tab:spec-freq} Pulsation frequencies determined from the radial velocity variations of the LSD~profiles, calculated using centroids.}
\centering
\begin{tabular}{lllllll}
\hline\hline
            & frequency    & amplitude  & phase       & $S/N$    \\
            & (d$^{-1}$)   & ($\rm km\,s^{-1}$)     & (rad)       &        \\
\hline
 $\nu_{a}$  & $2.24296(5)$ & $1.50(7)$  & $2.88(5)$   & $6.8$  \\
 $\nu_{b}$  & $2.35564(3)$ & $2.34(7)$  & $1.32(3)$   & $8.3$  \\
 $\nu_{c}$  & $2.46670(6)$ & $1.14(8)$  & $1.70(6)$   & $5.7$  \\
 $\nu_{d}$  & $2.51058(7)$ & $0.75(8)$  & $0.71(9)$   & $4.5$  \\
\hline
\end{tabular}
\end{table}

\section{The near-core rotation rate}
\label{sec:core-rot}
\subsection{Asymptotic asteroseismic modelling}
\label{subsec:asymptotic-modelling}
\begin{figure*}
    \centering
    \includegraphics[width=\textwidth]{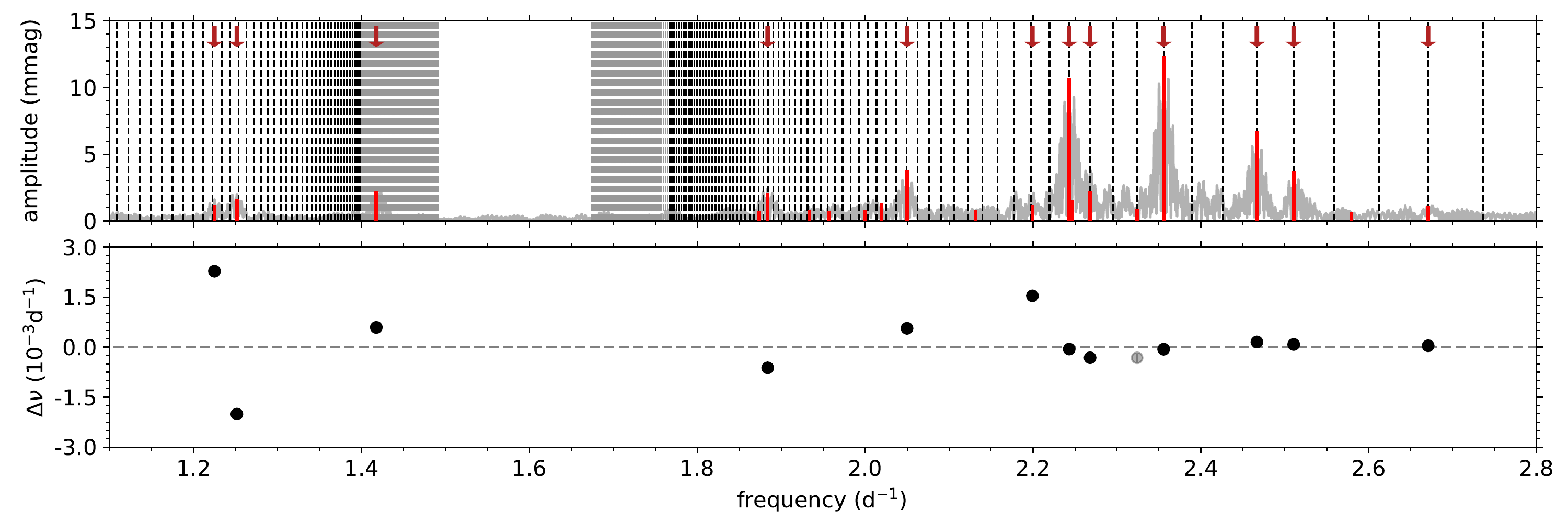}
    \caption{{\em Top:} Lomb-Scargle periodogram of the complete light curve (light grey) with the prewhitened pulsation frequencies with $S/N \geq 4.0$ (full red lines) and dark red arrows marking the frequencies with $S/N \geq 5.6$. The best-fitting asymptotic $g$-mode patterns (dashed lines) for the r~modes (left) and the prograde dipole modes (right) are also shown. The parts of the theoretical patterns that cannot be resolved properly (where the spacing $\Delta \nu$ between consecutive frequencies is less than twice the frequency resolution $\nu_{\rm res}$ are shown by the grey dashed lines. {\em Bottom:} the residuals (black) of the modelled observed frequencies, and the differences (grey) between the remaining observed frequencies and the closest model frequencies. $G$~modes with $4.0 \leq S/N < 5.6$ that are not shown in the bottom panel, lie outside of the plotted range.}
    \label{fig:best-pattern}
\end{figure*}

Next, we detected period-spacing patterns in the data and modelled them to measure the near-core rotation rate $\rm \nu_{rot}$ of HD\,112429. In the framework of the traditional approximation of rotation \citep[TAR; e.g.][]{Eckart1960,Bildsten1996,Lee1997}, where the horizontal component of the rotation vector is ignored, and in the asymptotic regime, where the pulsation frequencies in the co-rotating frame $\nu_{\rm co}$ are much smaller than the Brunt-V\"ais\"al\"a frequency $N/2\pi$ \citep[e.g.][]{Bouabid2013}, we have
\begin{equation}
    \nu_{nkm} = \left(\frac{\Pi_0}{\sqrt{\Lambda_{skm}}}\left(n + \alpha_g\right)\right)^{-1} + m\nu_{\rm rot}, \label{eq:per-sp}
\end{equation}
where $n$ is the radial order, $(k,m)$ is the pulsation mode identification \citep[e.g.][]{Lee1997}, $\Lambda_{skm}$ is the eigenvalue of the Laplace Tidal Equation, $\Pi_0$ is the buoyancy travel time given by
\begin{equation*}
    \Pi_0 = 2\pi^2\left(\int_{r_1}^{r_2}\frac{N(r)}{r}\mathrm{d}r\right)^{-1},
\end{equation*}
and $\alpha_g$ is a phase term, dependent on the mode behaviour at the boundaries ($r_1$,$r_2$) of the pulsation mode cavity. In most previous works, analysed stars with $g$~modes had been observed near-continuously for long time spans, such as stars in the original {\em Kepler\hyphenation{Kep-ler}} field-of-view \citep[e.g.][]{VanReeth2016,Li2020} or the TESS CVZs \cite[e.g.][]{Garcia2022}. 
By contrast, our TESS light curve of HD\,112429 exhibits large gaps, which resulted in a complex spectral window in the Fourier domain, as shown in Fig.\,\ref{fig:puls-freq}. This limits the number of measurable $g$-mode pulsation frequencies and has resulted in the possible inclusion of aliasing frequencies in our list. Hence, we took the following approach to measure the near-core $\nu_{\rm rot}$.
\begin{enumerate} 
\item[(1)] We modified the methodology developed by \citet{Christophe2018} to account for the sparse sampling of the $g$~modes. In summary, we selected a mode identification $(k,m)$ and evaluated a regularly sampled grid of rotation frequencies $\rm \nu_{rot}$, and used Eq.\,(\ref{eq:per-sp}) to rewrite the observed frequencies $\nu_{\rm obs,i}$ as 
\begin{equation*}
\frac{\sqrt{\Lambda_{skm}}}{\nu_{\rm obs,i} - m\nu_{\rm rot}} = \Pi_0\left(n_i + \alpha_g\right).
\end{equation*}
This is a linear relation as a function of $n_i$ and $\alpha_g$. We then assumed different values for $\Pi_0$ between 2300 and 5600\,s \citep{VanReeth2016}, and fit a linear model to estimate the radial orders $n_i$ and phase term $\alpha_g$ for each of them, starting out from the median period in the pattern and counting outwards. The quality of the model was then evaluated by fitting the theoretical pulsation frequencies (obtained from Eq.\,(\ref{eq:per-sp})) to the observed frequencies $\nu_{\rm obs,i}$ using a $\chi^2$-statistic.
\item[(2)] We limited this numerical analysis to the independent pulsation frequencies with $S/N \geq 5.6$, and visually compared the independent frequencies with $4 \leq S/N \leq 5.6$ with the predictions from the best-fitting asymptotic patterns.
\item[(3)] We determined the frequency spacings $\Delta \nu$ between consecutive theoretical frequencies in the best-fitting models, and excluded the observed frequencies for which the corresponding $\Delta \nu$ values are smaller than twice the frequency resolution $\nu_{\rm res}$, to ensure that the modelled observed frequencies are well-resolved. For our study of HD\,112429, only one frequency ($\nu_{11}$) had to be excluded during this step. We then repeated the asymptotic modelling of the remaining observed frequencies.
\item[(4)] To evaluate the impact of the sparse observed TESS sectors, we repeated and compared our analysis for the spectroscopic data and for different parts of the light curve: TESS sector 14, sectors 14-15, sectors 14-15 and 21-22, and sectors 14-15, 21-22 and 41. In each case, we applied iterative prewhitening, identified the significant independent $g$-mode frequencies and derived the radial order $n_i$, the mode identification $(k,m)$, the rotation rate $\nu_{\rm rot}$ and buoyancy travel time $\Pi_0$. The phase term $\alpha_g$, while a free parameter in our model, could not be measured precisely, and is not explicitly included in the subsequent discussions.
\end{enumerate}
The results of our analysis are listed in Table\,\ref{tab:frot-pi0} and illustrated in Fig.\,\ref{fig:best-pattern} for the best-fitting model pattern. When only TESS sector 14 was used, we only had a single $g$-mode measurement with $S/N \geq 5.6$, causing the $\nu_{\rm rot}$ measurement to fail. When two or more TESS sectors or the spectroscopic data were analysed, we obtained solutions for $\rm \nu_{rot}$ and $\Pi_0$ that are consistent with each other within the 95\,\% confidence interval, but not within 1-$\sigma$, as can be seen in Fig.\,\ref{fig:chi2}. It is well-known that the intrinsic structure of $g$-mode period-spacing patterns can bias the $\Pi_0$-$\nu_{\rm rot}$ measurements \citep[e.g.][]{VanReeth2016,Christophe2018,Takata2020}. In most previous studies, which relied on {\em Kepler} photometry, the solution was to exclude the largest outliers from the $g$-mode pattern. However, because the number of frequencies measured for HD\,112429 is limited, accurately identifying the true outliers was not possible. Instead we followed the same conservative approach used by \citet{Takata2020} and reported the $99\%$-confidence intervals in Table\,\ref{tab:frot-pi0}, to account for the sampling bias.

\begin{table}
\caption{\label{tab:frot-pi0} Measured $\rm \nu_{rot}$ and $\Pi_0$ values for the different data sets, with the 99\%-confidence intervals.}
\centering
\begin{tabular}{lll}
\hline\hline
data set           & $\nu_{rot}$          & $\Pi_{0}$ \\
                   & ($\rm d^{-1}$)      & (s)      \\
\hline
s14-15             & 1.5305\,(39)        & 4138\,(47) \\
s14-15, 21-22      & 1.5368\,(23)        & 4200\,(37) \\
s14-15, 21-22, 41  & 1.5360\,(33)        & 4190\,(49) \\
spectroscopy       & 1.5331\,(24)        & 4166\,(24) \\
\hline
\end{tabular}
\end{table}

\subsection{Conditions for $\nu_{\rm rot}$-$\Pi_0$ measurement}
\label{subsec:data-req}
\begin{figure}
    \centering
    \includegraphics[width=88mm]{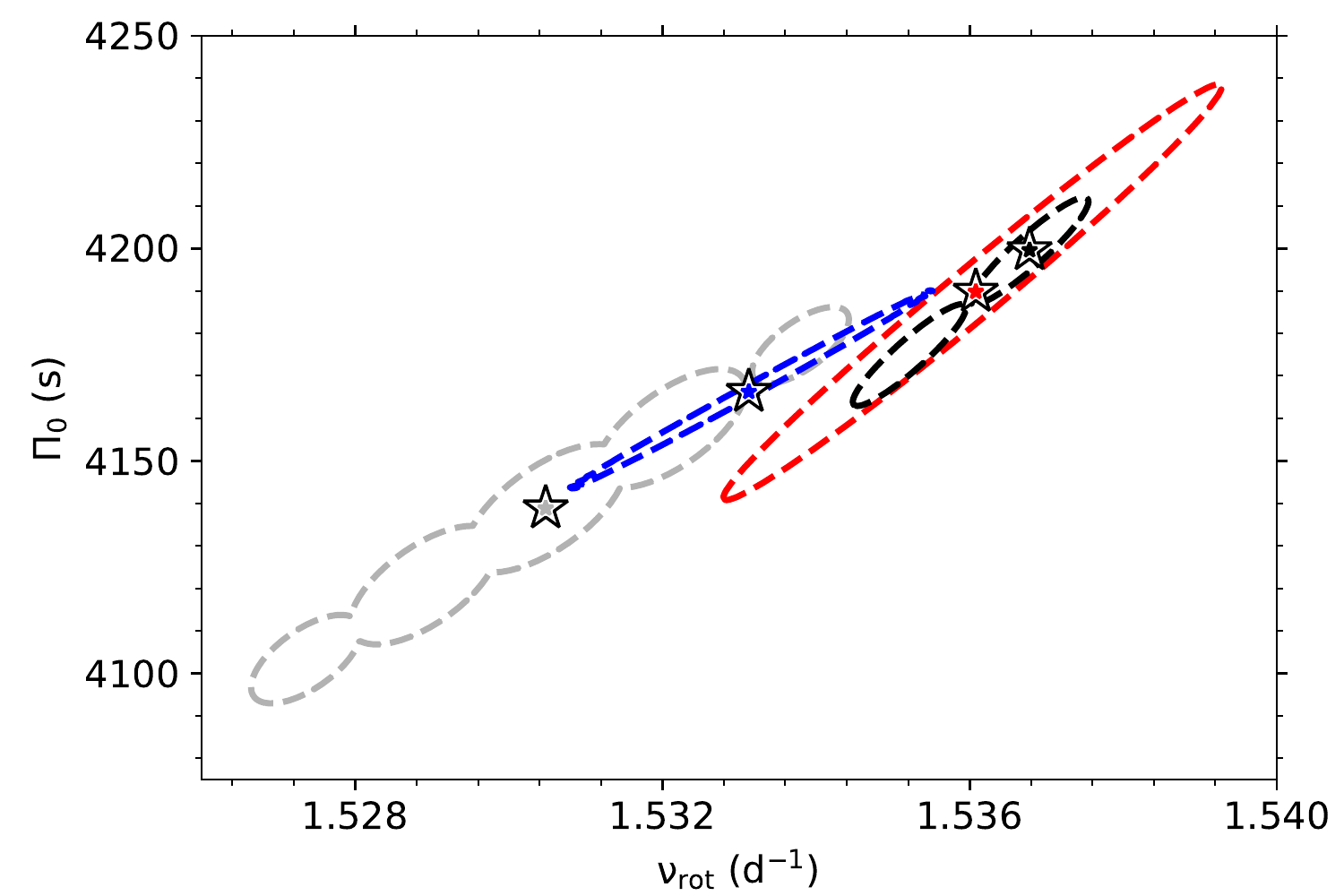}
    \caption{Confidence intervals for the best-fitting models of the pulsation frequencies, measured from TESS sectors 14-15 (light grey), sectors 14-15, 21-22 (black), sectors 14-15, 21-22 and 41 (red), and the spectroscopic data (blue). Confidence intervals at 3-$\sigma$ are indicated by dashed lines, while the stars mark the best-fitting models.}
    \label{fig:chi2}
\end{figure}

Most previous studies where $g$-mode patterns were detected relied on {\em Kepler} photometry. Because of the long time span (4 years) and high precision, most frequencies were often measured with error margins down to $10^{-6}$-$10^{-7}$\,$\rm d^{-1}$. Hence, it is an interesting result that we were able to detect the $g$-mode pattern and measure $\nu_{\rm rot}$ and $\Pi_0$ for HD\,112429, even when only two consecutive TESS sectors (spanning 54 days) were used (as shown in Fig.\,\ref{fig:chi2}). There are several factors that play a role here.

First, the radial orders of the observed $g$~modes of HD\,112429 are lower than those of the average $\gamma$\,Dor star observed with {\em Kepler}: based on the asymptotic modelling, the radial orders of the observed pulsations vary between $-19$ and $-60$, whereas the $\gamma$\,Dor stars in the sample of \citet{Li2020} have radial orders between $-10$ and $-100$, with an average of roughly $-50$. As a result, the frequency spacings between consecutive radial orders are larger for HD\,112429 than for many other $\gamma$\,Dor stars.

Second, the radial order spacing between the dominant $g$~modes allowed us to easily assign the correct radial orders to the observed pulsations: for the best-fitting model, the four dominant pulsations have radial orders $n$ equal to $-22$, $-23$, $-26$ and $-30$. The number of missing radial orders between them are multiples of different prime numbers: $\Delta n$ = 1, 3 and 4, respectively. This means it is easier to converge towards a unique solution of the radial-order identification of the different pulsations.

\begin{figure}
    \centering
    \includegraphics[width=88mm]{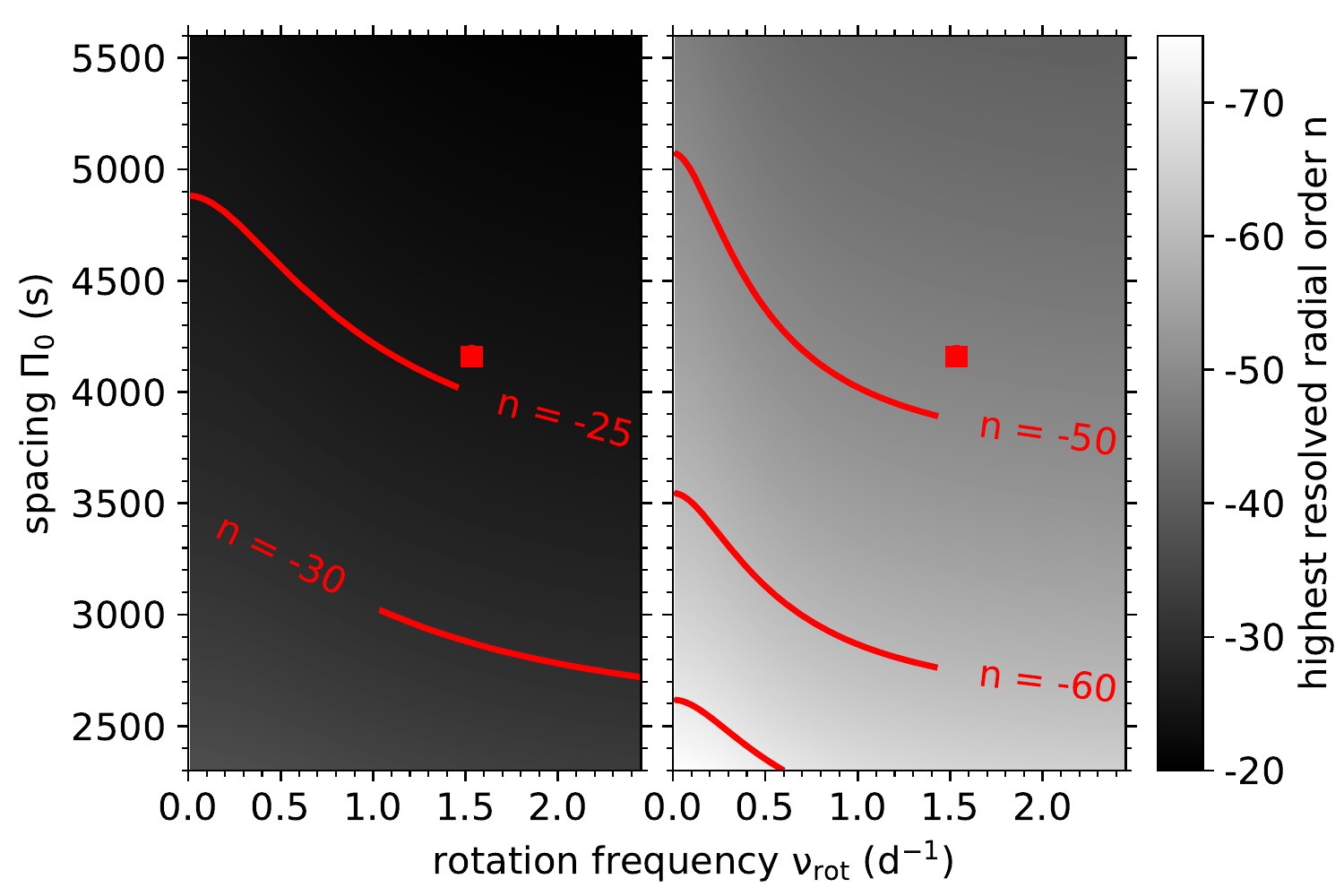}
    \caption{Highest radial orders of pulsation modes with $(k,m) = (0,1)$ that can theoretically be resolved with a light curve spanning two consecutive TESS sectors {\em (left)} or nine TESS sector {\em (right)}, as a function of the stellar buoyancy travel time $\Pi_0$ and rotation frequency $\nu_{\rm rot}$. It is assumed that the $(k,m) = (0,1)$ modes do not overlap with any other pulsations, and the effects of aliasing are ignored. The location of HD\,112429 is marked by a red square in both panels.}
    \label{fig:T2resolve}
\end{figure}

Third, the dominant $g$-mode pulsations that are located in the densest part of the pulsation frequency spectrum, do not have consecutive radial orders. In order to clearly resolve two consecutive pulsation frequencies, the spacing $\Delta \nu$ between them must be larger than twice the frequency resolution \citep{Christensen-Dalsgaard-Gough1982}. As illustrated in Fig.\,\ref{fig:T2resolve}, two TESS sectors are insufficient to properly resolve two of the four dominant $g$~modes from the pulsations with consecutive radial orders. However, these neighbouring modes have much lower amplitudes, so that their influence on the measured frequencies of the dominant $g$~modes is minimal. 

These three factors, combined with the methodology outlined above, allowed us to overcome the limitations of the data and measure $\nu_{\rm rot}$ and $\Pi_0$. Because of the large gaps in the TESS light curve, measured frequencies with lower $S/N$ are more likely to be affected by aliasing. This is probably why many of the pulsation frequencies with $S/N \leq 5.6$ (indicated in the top panel of Fig.\,\ref{fig:best-pattern}) do not match the theoretical asymptotic $g$-mode pattern (causing them to fall outside of the axis range of the bottom panel of Fig.\,\ref{fig:best-pattern}). Additionally, when fewer frequencies are measured and modelled, it is harder to evaluate if the model fit to the observed pattern is biased by the selection of observed frequencies. An extreme example of this effect are the spectroscopic data analysed in this work: we could only determine four frequencies from the spectroscopic time series, the absolute minimum to fit the pulsation model pattern with three free parameters ($\nu_{\rm rot}$, $\Pi_0$ and $\alpha_g$).

\section{Line profile variability}
\label{sec:lpv}
After the asymptotic frequency analysis, we verified the derived mode identification by evaluating the spectroscopic line profile variations (LPV) using the pixel-by-pixel method \citep{Zima2006}. First, we mapped the LSD~profiles from the different spectroscopic observations on a common velocity scale. We then determined the amplitude and phase profiles of the $g$-mode pulsations by fitting sine waves to the LSD~profiles at each radial velocity using the \texttt{lmfit} python package. To maximise the $S/N$ value of the results, we limited ourselves to the three dominant $g$-mode pulsations and excluded the data from the McKellar, SES and TCES spectrographs from this analysis. Despite the additional corrections that were done during the spectroscopic data reduction in Sect.~\ref{subsec:spectroscopy}, the remaining instrumental trends from these three instruments still influenced the data analysis results. 

\begin{figure*}
    \centering
    \includegraphics[width=\textwidth]{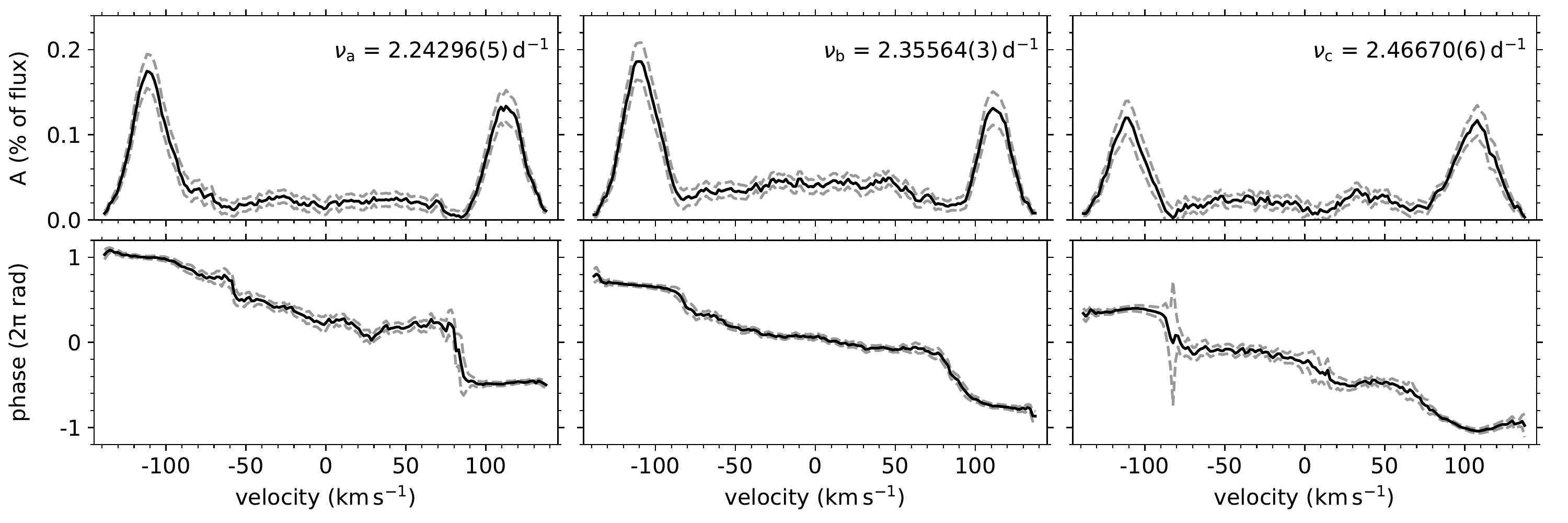}
    \caption{Observed line profile variations for the three dominant $g$-mode pulsations of HD\,112429 (black) with their 1-$\sigma$ error margins (grey dashed lines), calculated using the combined HERMES, HIDES and SOPHIE spectra, and displayed from left to right. The three amplitudes and phases of the line profile variations are shown in the top and bottom panels, respectively.}
    \label{fig:obs-lpv}
\end{figure*}

The final calculated amplitude and phase profiles of the $g$-mode LPV are shown in Fig.\,\ref{fig:obs-lpv}. Here, we can clearly see how small the pulsation amplitudes in the spectroscopic data are. The highest amplitudes, $\sim0.2\,\%$ of the normalised flux, are reached in the wings of the spectral lines. Near the centres of the profiles, the amplitudes have values between $0.01\,\%$ to $0.05\,\%$ of the normalised flux, or about one to five times their uncertainty. This explains why the residual instrumental variability from the McKellar, SES and TCES spectrographs had a big impact on the measured profiles, and those data had to be excluded.
 
\begin{figure}
    \centering
    \includegraphics[width=88mm]{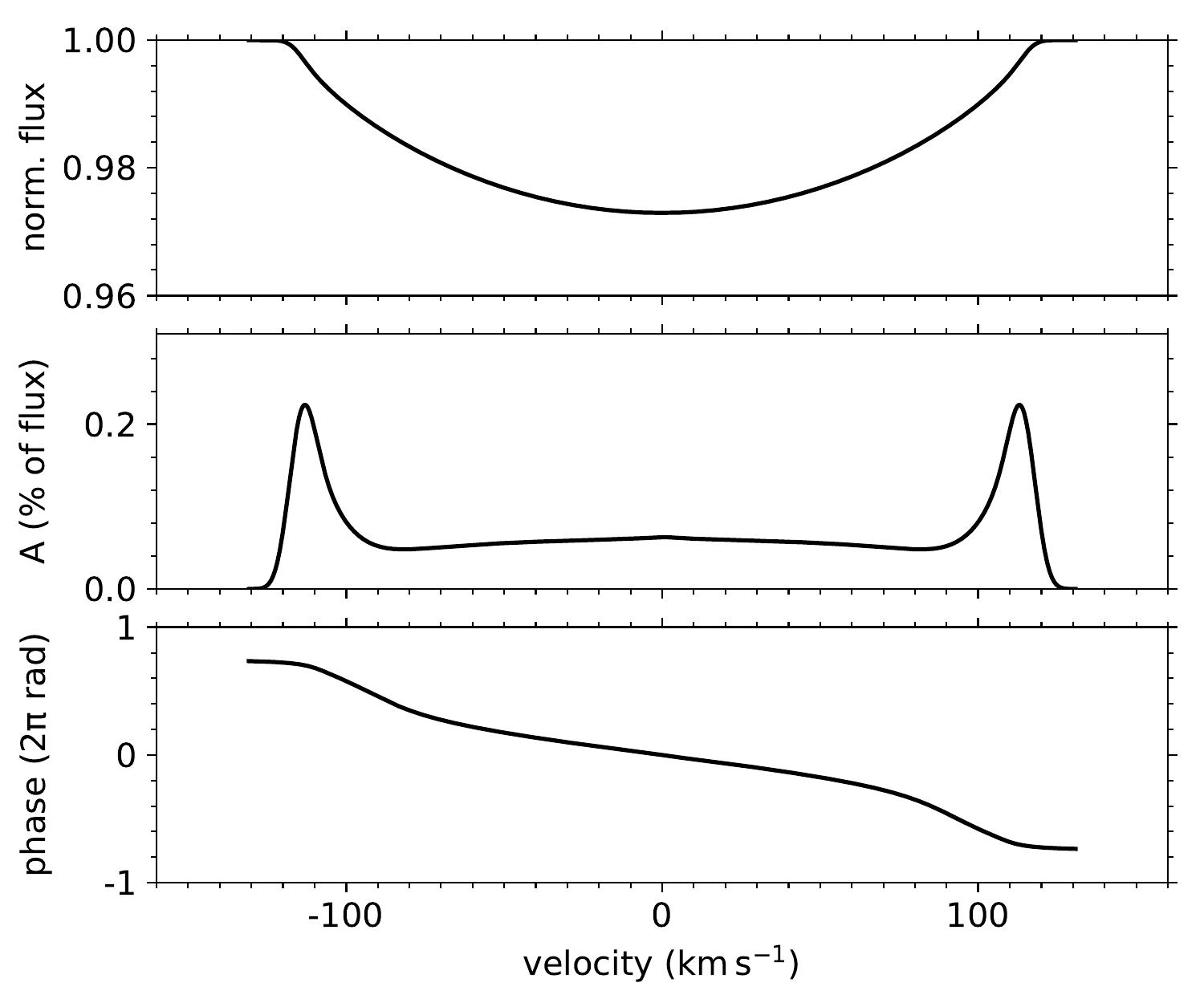}
    \caption{Theoretical line profile variations for a $g$~mode with $(n,k,m) = (-30,0,1)$ in a 1.5-M${}_\odot$ stellar model with $X_c = 0.4$, $\nu_{\rm rot} = 1.5\,\rm d^{-1}$ and an inclination angle $i = 60$\textdegree, calculated using the toy model described in Appendix\,\ref{app:gmode-lpv}. The velocity and temperature amplitudes of the pulsation at the stellar surface are set at $5\rm\,km\,s^{-1}$ and 10\,K, respectively.}
    \label{fig:best-theor-lpv}
\end{figure}

In the case of a sufficiently slowly rotating star, the theoretical model implemented in the FAMIAS program \citep{Zima2008} can be used to model observed LPV \citep[e.g.][]{Shutt2021}. However, because HD\,112429 is a moderate- to fast-rotating star, we instead built a toy model of the LPV within the TAR framework, as outlined in Appendix\,\ref{app:gmode-lpv}. To help calibrate the LPV model, we used theoretical $g$-mode pulsations, calculated with GYRE v6.0.1\footnote{ \href{https://gyre.readthedocs.io/}{https://gyre.readthedocs.io/}} \citep{Townsend2013,Townsend2018,Goldstein2020} for a stellar structure model of 1.5\,M$\rm {}_\odot$, with a core hydrogen fraction $X_{\rm c}$ of 0.4, solar metallicity, a core overshooting $f_{\rm ov}$ of 0.075 and extra diffusive mixing $D_{\rm mix}$ of 1\,$\rm cm^2\,s^{-1}$. This stellar model, calculated with the MESA code\footnote{\href{https://docs.mesastar.org/}{https://docs.mesastar.org/}} v11701 \citep[e.g.][]{Paxton2011,Paxton2019}, is in rough agreement with the parameters of HD\,112429 presented in Table\,\ref{tab:hd112429} and the buoyancy travel time derived in Sect.\,\ref{sec:core-rot}. The resulting theoretical LPV are shown in Fig.\,\ref{fig:best-theor-lpv} for a $g$-mode pulsation with $(n,k,m) = (-30,0,1)$, an assumed rotation rate of $1.5\,\rm d^{-1}$ and an inclination angle $i$ of 60\textdegree. The velocity and temperature amplitudes of the pulsation at the stellar surface were set at $5\rm\,km\,s^{-1}$ and 10\,K, respectively. In short, if a $g$-mode pulsation is in the subinertial regime, where spin $s = 2\nu_{\rm rot}/\nu_{\rm co} > 1$, it is confined within a band around the equator by the Coriolis force. Moreover, in the case of $(k,m) = (0,1)$ $g$~modes, the dominant component of the pulsation displacement is also parallel to the equator. As a result, the LPV are strongest in the wings of the spectral lines. This is in agreement with the observed LPV shown in Fig.\,\ref{fig:obs-lpv}, confirming the results we obtained in Sect.~\ref{sec:core-rot}.

\section{Discussion and conclusions}
\label{sec:discussion}
HD\,112429 is a known single $\gamma$\,Dor star, first detected by \citet{Aerts1998}, for which two years of legacy ground-based spectroscopy from six different spectrographs are available, as outlined by \citet{DeCat2009}. We combined these data with new photometric observations from the TESS space mission, which also span over two years, for an in-depth analysis of the star.

Following a custom data reduction of the TESS photometry, we measured and analysed the $g$-mode pulsation frequencies, and used them to determine the near-core rotation rate $\nu_{\rm rot}$ and buoyancy travel time $\Pi_0$, with resulting values of $1.536\,(3)\,\rm d^{-1}$ and $4190\,(50)\,$s, respectively, where the provided error margins correspond to the $99\%$ confidence interval. Since there are only five sectors of TESS data available and there are large gaps in the light curve, we repeated our analysis for subsections of the light curves to establish the effect of the sampling on our analysis results. We demonstrated that at the 1-$\sigma$ level, there are indeed systematic offsets on the derived solutions. This is in agreement with  earlier studies in the literature, which demonstrated that the intrinsic structure of $g$-mode patterns can bias the $\nu_{\rm rot}$ and $\Pi_0$ measurements. Because the number of frequencies detected for HD\,112429 is limited, our results are more sensitive to this bias. However, the parameter values we derived for the different data sets do agree with each other within the $99\%$ confidence interval, the same conservative range employed by \citet{Takata2020}.

The results from the asymptotic modelling of the space photometry was subsequently confirmed by a pixel-by-pixel evaluation of the LSD profile variations, which were calculated for 509 spectra taken with the HERMES, HIDES and SOPHIE spectrographs. We measured the LPV for the three dominant $g$-mode pulsations, and to account for the moderate- to fast-rotation rate of HD\,112429, we developed a toy model for the LPV within the TAR framework. The observed LPV were qualitatively similar to our simulations, computed for $(k,m) = (0,1)$ $g$~modes in a 1.5-M${}_\odot$ stellar model rotating at $1.5\,\rm d^{-1}$, seen at an angle of $\sim$60\textdegree. In both cases, the dominant variability is located in the wings of the spectral line profiles, because of the equatorial confinement of the $g$~modes by the Coriolis force. In the future, a more detailed evaluation of the observed LPV can provide us with more accurate constraints on the displacement and velocity fields of $g$-mode pulsations in fast-rotating stars. It is interesting to see how the observed LPV are even more strongly confined to the wings of the lines than the variability in our toy model. However, because of the low amplitudes of the observed pulsations and the required high $S/N$ of the data, the amount of information that can be gained by studying a spectroscopic time series, is limited compared to what can be learned from space photometry, especially when observations from different spectrographs with significant offsets between their calibrations are combined. The main advantage of using ground-based data from different observation sites, that is reducing the one-day aliasing caused by the Earth's rotation, is less important when space-based photometry is available, because these data suffer much less from aliasing. For most fast-rotating $g$-mode pulsators, a detailed analysis of space photometry, supported by a precise measurement of atmospheric parameters such as from \citet[][see Table \ref{tab:hd112429}]{Kahraman2016}, will be more instructive.

Almost all previous studies in which near-core rotation rates of $g$-mode pulsators were measured, relied on long-timebase space photometry from {\em Kepler} \citep[e.g.][]{VanReeth2016,Li2020,Christophe2018} or the TESS CVZs \citep{Garcia2022}. The resulting accurate $g$-mode frequency measurements allow for precise $\nu_{\rm rot}$ and $\Pi_0$ determination. Other studies in the literature that relied on shorter data sets for such analyses \citep[e.g.][]{Zwintz2017,Christophe2020} typically resulted in much larger error margins for the measured $\nu_{\rm rot}$ and $\Pi_0$ values. Hence, our precise parameter determination for HD\,112429 with no more than two TESS sectors raises the interesting question what the minimum data requirements for asymptotic $g$-mode modelling are and how these depend on the stellar properties. Long-timebase observations are of course preferable for detailed $g$-mode analysis, since they can reveal more precise information about the interior stellar structure \citep[e.g.][]{Bowman2021,Mombarg2021,Pedersen2021}. However, with the ongoing TESS space mission and the future PLATO mission providing shorter light curves for most stars on the sky, the ability to constrain the stellar properties from such data would allow us to do a global characterisation of a subset of $g$-mode pulsators across the entire sky, rather than only those that have been observed for years.

\begin{acknowledgements}
We thank the referee for their positive report and useful comments which improved the quality of the paper. TVR gratefully acknowledges a postdoctoral fellowship from the Research Foundation Flanders (FWO) with grant agreement N\textdegree 12ZB620N. JVB acknowledges receiving support from the Research Foundation Flanders (FWO) under grant agreement N\textdegree V421221N. The research leading to these results received partial funding from the KU Leuven Research Council (grant C16/18/005: PARADISE, with PI Conny Aerts). ANC is supported by the international Gemini Observatory, a program of NSF’s NOIRLab, which is managed by the Association of Universities for Research in Astronomy (AURA) under a cooperative agreement with the National Science Foundation, on behalf of the Gemini partnership of Argentina, Brazil, Canada, Chile, the Republic of Korea, and the United States of America.

This paper includes data collected with the TESS mission, obtained from the MAST data archive at the Space Telescope Science Institute (STScI). Funding for the TESS mission is provided by NASA's Explorer Program. We thank the whole team for the development and operations of the mission. STScI is operated by the Association of Universities for Research in Astronomy, Inc., under NASA contract NAS 5-26555. We are also very grateful to Bill Paxton, Rich Townsend and their respective teams for the development and maintenance of the open-source stellar evolution code MESA and the open-source stellar pulsation code GYRE. We would also like to thank Christoffel Waelkens and Roy {\O}stensen for their contributions to the multi-site spectroscopic observation campaign, and Andrew Tkachenko for useful discussions and for his work on the code for least-squares deconvolution profile computation.

This work was also based on observations made with multiple spectrographs, including the SOPHIE spectrograph at the 1.93-m telescope at Observatoire de Haute-Provence (CNRS/OAMP), France; the McKellar spectrograph at the 1.2-m Telescope at the Dominion Astrophysical Observatory, NRC Herzberg, Programs in Astronomy and Astrophysics, National Research Council of Canada; the HERMES spectrograph, which is supported by the Research Foundation - Flanders (FWO), Belgium, the Research Council of KU Leuven, Belgium, the Fonds National de la Recherche Scientifique (F.R.S.-FNRS), Belgium, the Royal Observatory of Belgium, the Observatoire de Genève, Switzerland and the Th\"uringer Landessternwarte Tautenburg, Germany, and installed at the Mercator telescope, operated by the Flemish Community, on the island of La Palma at the Spanish Observatorio del Roque de los Muchachos of the Instituto de Astrof\'isica de Canarias (IAC); the HIDES spectrograph at the 1.88-m telescope at the Okayama Astrophysical Observatory in Japan; the Sandiford Cassegrain Echelle Spectrograph at the Otto Struve telescope at the McDonald Observatory of the University of Texas at Austin; the TCES spectrograph at the 2.0-m Alfred Jensch Telescope at the T\"uringer Landessternwarte in Tautenburg, Germany.

This research made use of the SIMBAD database, operated at CDS, Strasbourg, France, the SAO/NASA Astrophysics Data System, the VizieR catalogue access tool, CDS, Strasbourg, France, and Lightkurve, a Python package for Kepler and TESS data analysis (Lightkurve Collaboration, 2018).
.
\end{acknowledgements}

\bibliographystyle{aa}
\bibliography{HD112429}

\appendix
\section{g-mode line profile variations in moderate- to fast-rotating stars}
\label{app:gmode-lpv}
\begin{figure}
    \centering
    \includegraphics[width=88mm]{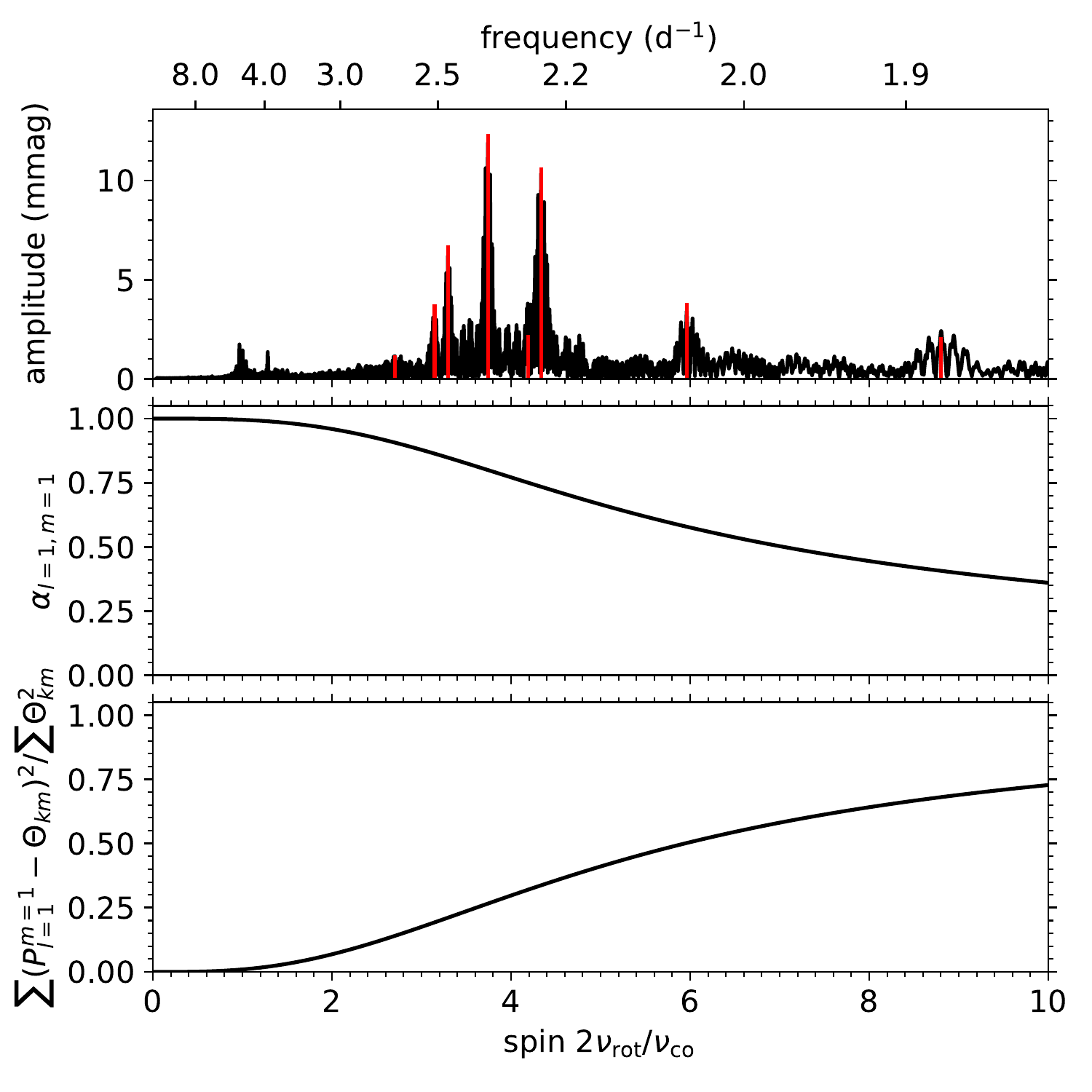}
    \caption{Validity evaluation of approximating $(k,m) = (0,1)$ Hough functions $\Theta_{km}(\theta,s)$ with a Legendre polynomial $P^{m=1}_{l=1}(\theta)$, as a function of the spin parameter. {\em Top:} Lomb-Scargle periodogram of HD\,112429, with the observed $(k,m) = (0,1)$ pulsations (red), as a function of the spin parameter. A comparison with the bottom two panels reveals the relative influence of the Coriolis force on each observed pulsation mode. {\em Middle:} relative contribution of the $P^{m=1}_{l=1}(\theta)$ function when the Hough function is expressed as a sum of Legendre polynomials $\Theta(\theta,s) = \sum \alpha_{lm}P^l_m(\theta)$. {\em Bottom:} the relative residual sum of squares, after fitting the $P^{m=1}_{l=1}(\theta)$ Legendre polynomial to the Hough function $\Theta_{km}$.}
    \label{fig:TAR-eval}
\end{figure}

In moderate- to fast-rotating stars, the Coriolis force confines $g$-mode pulsations to an equatorial band \citep[e.g.][]{Townsend2003,Saio2018}. Hence, while $g$~modes in non-rotating or slowly rotating stars can be described by spherical harmonics $Y^m_l(\theta,\phi) = P^m_l(\theta)\exp\left(im\phi\right)$, this is no longer the case for $g$-mode pulsations in rotating stars with spin $s = 2\nu_{\rm rot}/\nu_{\rm co} > 1$. Within the TAR framework, the latitudinal dependence of the pulsation eigenmodes is represented by Hough functions $\Theta_{km}(\theta,s)$, which in turn can be expressed as a sum of Legendre polynomials \citep[e.g.][]{Saio2018}: $$H_{km}(\theta,\phi,s) = \Theta_{km}(\theta,s)\exp\left(im\phi\right) = \sum_{l \geq |m|} \alpha_{l,m}(s)P^m_l(\theta)\exp\left(im\phi\right),$$ with real coefficients $\alpha_{l,m}(s)$. This is illustrated in Fig.\,\ref{fig:TAR-eval} for the Hough function $\Theta_{k=0,m=1}$. At spin values $s < 1$, the coefficient $\alpha_{l=1,m=1}$ of the Legendre polynomial $P^{m=1}_{l=1}$ is close to unity. At higher spin values, $\alpha_{l,m}$ rapidly decreases, as does the variance of $\Theta_{k=0,m=1}$ that is explained by $P^{m=1}_{l=1}$. When these graphs are compared to the spin values of the observed $(k,m) = (0,1)$-modes of HD\,112429, we see that the spherical harmonics $Y^{m=1}_{l=1}$ do not suffice here. 

\begin{figure}
    \centering
    \includegraphics[width=49mm]{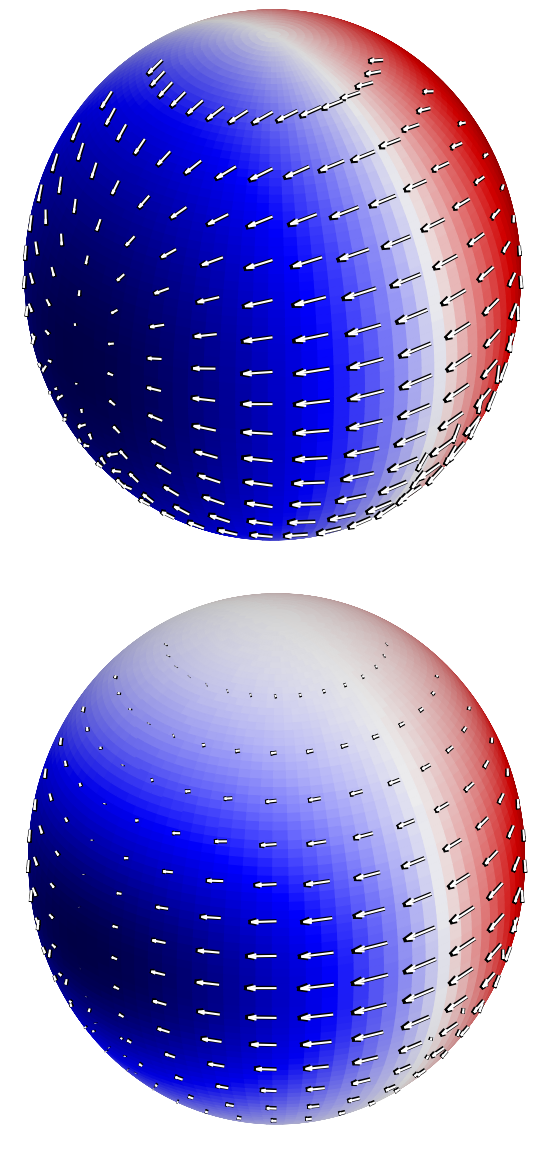}
    \caption{Temperature variations (blue = hot, red = cold) and displacements (white arrows) associated with a $(k,m) = (0,1)$ $g$~mode at spin values of 0.58 {\em (top)} and 4.15 {\em (bottom)}. These mode geometries match those of the $(n,k,m) = (-30,0,1)$ pulsation evaluated in Fig.\,\ref{fig:theor_LPV} for a 1.5-M${}_\odot$ stellar model with a core hydrogen fraction of $X_c = 0.4$, seen at an inclination angle of 60\textdegree\ and rotating at $0.25\,\rm d^{-1}$ {\em (top)} and $1.5\,\rm d^{-1}$ {\em (bottom)}, respectively.}
    \label{fig:TAR-eigenmodes}
\end{figure}

\begin{figure*}
    \centering
    \includegraphics[width=\textwidth]{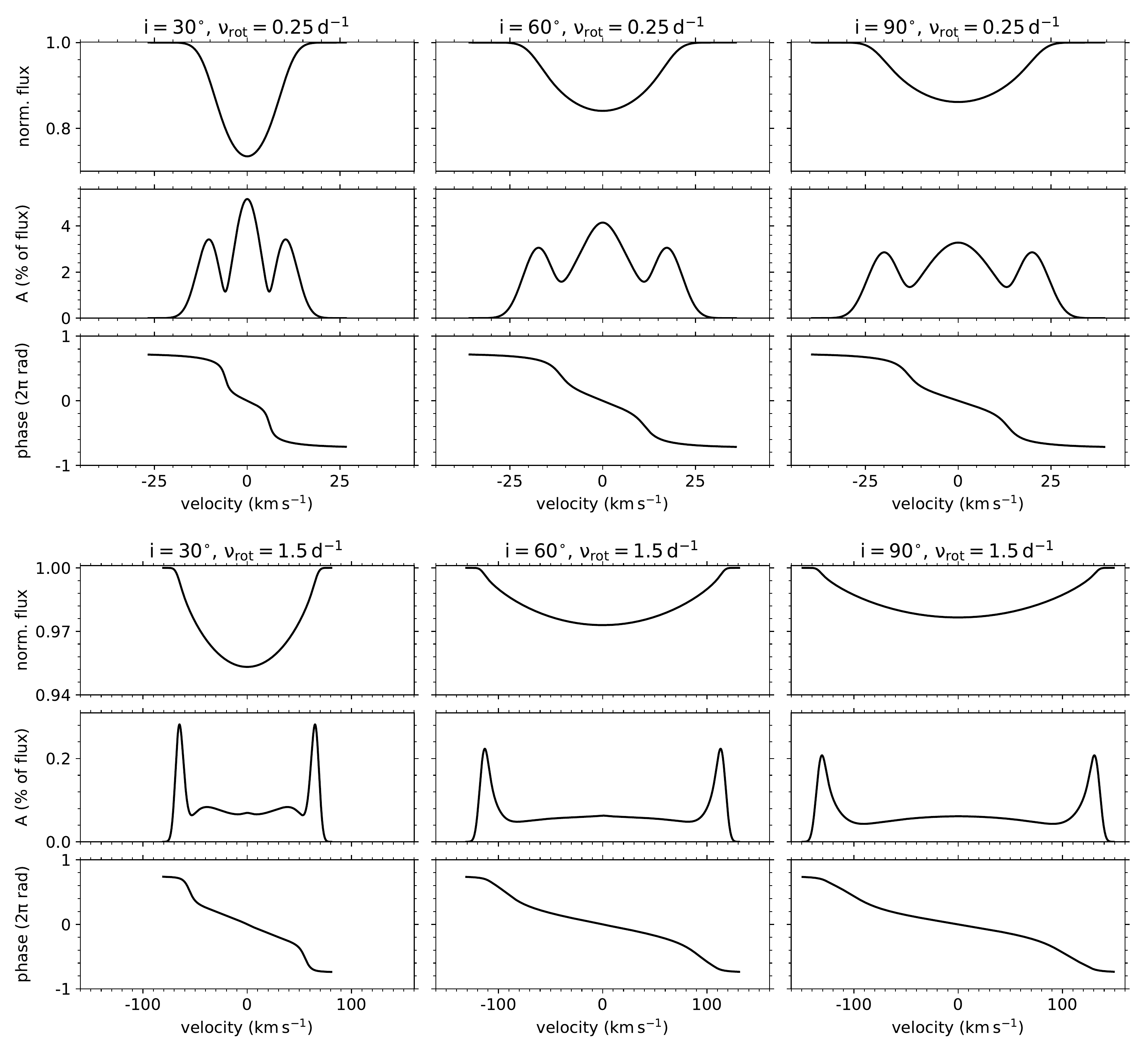}
    \caption{Theoretical LSD~profile variations for a $g$~mode with $(n,k,m) = (-30,0,1)$ in a 1.5-M${}_\odot$ stellar model with $X_c = 0.4$, calculated using a toy model (described in Sect.\,\ref{subsec:toymodel}). The velocity and temperature amplitudes of the pulsation at the stellar surface are set at $5\rm\,km\,s^{-1}$ and 10\,K, respectively. The LPV are shown for $\nu_{\rm rot}$ values of 0.25\,$\rm d^{-1}$ {(\em top)} and $1.5\,\rm d^{-1}$ {(\em bottom)}, at inclination angles of 30\textdegree {(\em left)}, 60\textdegree {(\em middle)} and 90\textdegree {(\em right)}.}
    \label{fig:theor_LPV}
\end{figure*}

This fundamentally changes the spectral line profile variations (LPV) caused by the observed $g$-mode pulsations. To evaluate this qualitatively, we built a toy model as explained below in Sect.\,\ref{subsec:toymodel}. The result is illustrated in Figs. \ref{fig:TAR-eigenmodes} and \ref{fig:theor_LPV} for a $g$~mode with $(n,k,m) = (-30,0,1)$, in a 1.5-M${}_\odot$ MESA model at $X_c = 0.4$ \citep[v11701; e.g.][]{Paxton2011,Paxton2019}. At low spin values ($s = 0.58$ in the top panels of Figs. \ref{fig:TAR-eigenmodes} and \ref{fig:theor_LPV}), the $g$~mode propagates at all colatitudes of the star, and LPV are detected in the whole LSD~profile. At moderate to high spin values ($s = 4.15$ in the bottom panels of Figs. \ref{fig:TAR-eigenmodes} and \ref{fig:theor_LPV}), the $g$~mode is confined in an equatorial band, leading to smaller LPV that are mostly detected in the wings of the LSD~profile. Qualitatively, the observed LPV of HD\,112429 shown in Fig.\,\ref{fig:obs-lpv} are very similar to the theoretical LPV of the toy model for $\nu_{\rm rot} = 1.5\,\rm d^{-1}$ at an inclination angle of $60$\textdegree, in agreement with the photometric analysis results in Sect.\,\ref{sec:core-rot}. 

\subsection{a toy model for $g$-mode LPV}
\label{subsec:toymodel}
In a spherically symmetric star with uniform rotation frequency $\nu_{\rm rot}$, the rotational velocity field at the stellar surface is given by
\begin{equation}
    v_{\rm rot}\left(\theta,\phi\right) = 2\pi\nu_{\rm rot}R_*\sin\theta,\label{eq:vrot}
\end{equation}
where $R_*$ is the stellar radius and $(\theta,\phi)$ are the angular spherical coordinates with respect to the rotation axis.

If we then apply the TAR to a $g$~mode with frequency $\nu_{\rm co}$ (in the co-rotating frame) and mode identification $(k,m)$ in this star, the displacement $\xi$ at the stellar surface can be expressed as \begin{align}
    &\xi_r\left(\theta,\phi\right) = -u_{sc}\mathfrak{R} \Theta_{km}\left(\theta, s\right)\cos\left(2\pi\nu_{\rm co}t + m\varphi\right)\\
    &\xi_\theta\left(\theta,\phi\right) = u_{sc} \Theta_{km}^{\theta}\left(\theta, s\right)\cos\left(2\pi\nu_{\rm co}t + m\varphi\right)\\
    &\xi_\phi\left(\theta,\phi\right) = -u_{sc} \Theta_{km}^{\phi}\left(\theta, s\right)\sin\left(2\pi\nu_{\rm co}t + m\varphi\right)
\end{align}
so that the pulsation velocity field is given by
\begin{align}
    &u_r\left(\theta,\phi\right) = 2\pi\nu_{\rm co}u_{sc}\mathfrak{R} \Theta_{km}\left(\theta, s\right)\sin\left(2\pi\nu_{\rm co}t + m\varphi\right)\label{eq:ur}\\
    &u_\theta\left(\theta,\phi\right) = -2\pi\nu_{\rm co}u_{sc} \Theta_{km}^{\theta}\left(\theta, s\right)\sin\left(2\pi\nu_{\rm co}t + m\varphi\right)\\
    &u_\phi\left(\theta,\phi\right) = -2\pi\nu_{\rm co}u_{sc} \Theta_{km}^{\phi}\left(\theta, s\right)\cos\left(2\pi\nu_{\rm co}t + m\varphi\right)\label{eq:up}
\end{align}
Here $\Theta_{km}$, $\Theta^{\theta}_{km}$ and $\Theta^{\phi}_{km}$ are the Hough functions associated with the mode identification $(k,m)$ at spin $s = 2\nu_{\rm rot}/\nu_{\rm co}$ \citep{Wang2016}. $\mathfrak{R}$ is a single real value, equal to the ratio of the amplitudes of $\xi_r$ to $\xi_h = \sqrt{\xi^2_\theta + \xi^2_\phi}$ at the stellar surface. That is, $\mathfrak{R} = |\xi_r| / \sqrt{|\xi_\theta|^2 + |\xi_\phi|^2}$. Finally, $u_{sc}$ is a user-defined scaling factor, such that the pulsation velocity amplitude $u$ can be set to a reasonable value. The pulsation temperature perturbation can be described by
\begin{equation}
    T'\left(\theta,\phi\right) = T_{sc} \Theta_{km}\left(\theta,s\right)\cos\left(2\pi\nu_{\rm co}t + m\varphi\right),
\end{equation}
with a user-defined scaling factor $T_{sc}$. An approximation of the local brightness variation $B\left(\theta,\phi\right)$ can then be calculated as \begin{equation}
    B\left(\theta,\phi\right) = \left(\frac{R_* + \xi_r\left(\theta,\phi\right)}{R_*}\right)^2\left(\frac{T_{\rm eff} + T'\left(\theta,\phi\right)}{T_{\rm eff}}\right)^4,\label{eq:brightvar}
\end{equation} where $T_{\rm eff}$ is the effective temperature of the stellar model. 

If the star is observed at an inclination angle $i$, we can define a new set of angular spherical coordinates $\left(\theta_L,\phi_L\right)$ with respect to the observer's line-of-sight, and calculate the stellar line profile $f(v)$ at a given time $t$ as
\begin{equation}
  \begin{split}
    f(v) = 1 - \int_0^{2\pi} \mathrm{d}\phi_L &\int_0^{\pi/2} \mathrm{d}\theta_L \sin\left(2\theta_L\right) \left[1 - \mu\left(1 - \cos\theta_L\right)\right ]\\
    &\times A \exp\left[\frac{-\left(v - v_L\left(\theta_L,\phi_L\right)\right)^2}{2v_{\rm therm}^2}\right]B\left(\theta,\phi\right),
  \end{split}
\end{equation}
where $\mu$ is the linear limb-darkening coefficient, $A$ is the spectral line depth in the non-rotating case, $v_{\rm therm}$ is the local thermal line broadening, $v_L\left(\theta_L,\phi_L\right)$ is the projection of the combined velocity field (from Eqs.(\ref{eq:vrot}) and (\ref{eq:ur}) to (\ref{eq:up})) along the observer's line-of-sight and $B\left(\theta,\phi\right)$ is obtained from Eq.\,(\ref{eq:brightvar}). In this work, $\mu$ and $v_{\rm therm}$ are set to 0.6 \citep[an approximate linear limb darkening coefficient value for $\gamma$\,Dor stars;][]{ClaretBloemen2011} and $3\,\rm km\,s^{-1}$ (as listed in Table\,\ref{tab:hd112429}), respectively. In the calculations for Fig.\,\ref{fig:theor_LPV}, the ratio $\mathfrak{R}$ was calculated with GYRE v6.0.1 \citep{Townsend2013,Townsend2018,Goldstein2020} for the used 1.5-M${}_\odot$ MESA model. A python implementation of the algorithm explained here, is available at \href{https://github.com/TVanReeth/LPV-simulations}{https://github.com/TVanReeth/LPV-simulations}.
\end{document}